\documentstyle[12pt]{article}

\let\useblackboard=\iftrue
\iftrue
\fi

\def\hybrid{\topmargin -20pt  \oddsidemargin 0pt
      \headheight 0pt   \headsep 0pt
      \textwidth 6.25in 
      \textheight 9.5in 
      \marginparwidth .875in
      \parskip 5pt plus 1pt   \jot = 1.5ex}

\hybrid

\let\LARGE=\large

\let\large=\normalsize

\begin{document}
\def\x{\times}
\def\beq{\begin{equation}}
\def\eeq{\end{equation}}
\def\beqa{\begin{eqnarray}}
\def\eeqa{\end{eqnarray}}
\def\L{ {\cal L}}
\def\C{ {\cal C}}
\def\N{ {\cal N}}
\def\calE{{\cal E}}
\def\lin{{\rm lin}}
\def\Tr{{\rm Tr}}
\def\cF{{\cal F}}
\def\cD{{\cal D}}
\def\modS{{S+\bar S}}
\def\mods{{s+\bar s}}
\newcommand{\Fg}[1]{{F}^{({#1})}}
\newcommand{\cFg}[1]{{\cal F}^{({#1})}}
\newcommand{\cFgc}[1]{{\cal F}^{({#1})\,{\rm cov}}}
\newcommand{\Fgc}[1]{{F}^{({#1})\,{\rm cov}}}
\def\mpl{m_{\rm Planck}}
\def\mxth{\mathsurround=0pt }
\def\xversim#1#2{\lower2.pt\vbox{\baselineskip0pt \lineskip-.5pt
x  \ialign{$\mxth#1\hfil##\hfil$\crcr#2\crcr\sim\crcr}}}
\def\simgr{\mathrel{\mathpalette\xversim >}}
\def\simle{\mathrel{\mathpalette\xversim <}}

\newcommand{\ms}[1]{\mbox{\scriptsize #1}}
\renewcommand{\a}{\alpha}
\renewcommand{\b}{\beta}
\renewcommand{\c}{\gamma}
\renewcommand{\d}{\delta}
\newcommand{\th}{\theta}
\newcommand{\TH}{\Theta}
\newcommand{\pa}{\partial}
\newcommand{\g}{\gamma}
\newcommand{\G}{\Gamma}
\newcommand{\A}{\Alpha}
\newcommand{\B}{\Beta}
\newcommand{\D}{\Delta}
\newcommand{\e}{\epsilon}
\newcommand{\E}{\Epsilon}
\newcommand{\z}{\zeta}
\newcommand{\Z}{\Zeta}
\newcommand{\k}{\kappa}
\newcommand{\K}{\Kappa}
\renewcommand{\l}{\lambda}
\renewcommand{\L}{\Lambda}
\newcommand{\m}{\mu}
\newcommand{\M}{\Mu}
\newcommand{\n}{\nu}
\newcommand{\X}{\Chi}
\newcommand{\R}{\Rho}
\newcommand{\s}{\sigma}
\renewcommand{\S}{\Sigma}
\renewcommand{\t}{\tau}
\newcommand{\T}{\Tau}
\newcommand{\y}{\upsilon}
\newcommand{\Y}{\upsilon}
\renewcommand{\o}{\omega}
\newcommand{\q}{\theta}
\newcommand{\h}{\eta}

\def\dota{ {\dot{\alpha}} }
\def\lag{Lagrangian}
\def\Kahler{K\"{a}hler}
\def\kahler{K\"{a}hler}
\def\A{ {\cal A}}
\def\C{ {\cal C}}
\def\F{{\cal F}}
\def\cL{ {\cal L}}

\def\R{ {\cal R}}
\def\x{ \times }
\def\beq{\begin{equation}}
\def\eeq{\end{equation}}
\def\beqa{\begin{eqnarray}}
\def\eeqa{\end{eqnarray}}

\sloppy
\newcommand{\ba}{\begin{array}}
\newcommand{\ea}{\end{array}}
\newcommand{\eq}{\begin{equation}}

\newcommand{\ov}{\overline}
\newcommand{\un}{\underline}
\newcommand{\p}{\partial}
\newcommand{\la}{\langle}
\newcommand{\ra}{\rangle}
\newcommand{\bl}{\boldmath}
\newcommand{\ds}{\displaystyle}
\newcommand{\nl}{\newline}
\newcommand{\Nzahl}{{\bf N}  }
\newcommand{\zzahl}{ {\bf Z} }
\newcommand{\Zzahl}{ {\bf Z} }
\newcommand{\Qzahl}{ {\bf Q}  }
\newcommand{\Rzahl}{ {\bf R} }
\newcommand{\Czahl}{ {\bf C} }
\newcommand{\wt}{\widetilde}
\newcommand{\wh}{\widehat}
\newcommand{\fs}[1]{\mbox{\scriptsize \bf #1}}
\newcommand{\ft}[1]{\mbox{\tiny \bf #1}}
\newtheorem{satz}{Satz}[section]
\newenvironment{Satz}{\begin{satz} \sf}{\end{satz}}
\newtheorem{definition}{Definition}[section]
\newenvironment{Definition}{\begin{definition} \rm}{\end{definition}}
\newtheorem{bem}{Bemerkung}
\newenvironment{Bem}{\begin{bem} \rm}{\end{bem}}
\newtheorem{bsp}{Beispiel}
\newenvironment{Bsp}{\begin{bsp} \rm}{\end{bsp}}


\renewcommand{\thesection}{\arabic{section}}
\renewcommand{\theequation}{\thesection.\arabic{equation}}

\parindent0em

\useblackboard
\typeout{If you do not have msbm (blackboard bold) fonts,}
\typeout{change the option at the top of the tex file.}
\font\blackboard=msbm10 scaled \magstep1
\font\blackboards=msbm7
\font\blackboardss=msbm5
\newfam\black
\textfont\black=\blackboard
\scriptfont\black=\blackboards
\scriptscriptfont\black=\blackboardss
\def\Bbb#1{{\fam\black\relax#1}}
\else
\def\Bbb{\bf}
\fi


\newcommand{\eqn}{\begin{eqnarray}}
\newcommand{\enq}{\end{eqnarray}}
\newcommand{\eqa}{\begin{array}}
\newcommand{\ena}{\end{array}}
\newcommand{\en}{\end{equation}}
\newcommand{\ie}{{\it i.e.}}
\newcommand{\no}{\nonumber}
\newcommand{\noi}{\noindent}
\newcommand{\ZZ}{Z\!\!\!Z}
\def\kite#1#2{$^{\mbox{\footnotesize \rm   #1-#2 \normalsize}}$}
\def\skite#1{$^{\mbox{\footnotesize \rm   #1 \normalsize}}$}
\def\comments#1{}
\newcommand{\IZ}{{\Bbb{Z}}}
\def\N{N}
\def\1N{$1\over N$}
\def\CC{\Bbb{C}}
\def\RR{\Bbb{R}}
\def\IZ{\Bbb{Z}}
\def\del{\partial}
\def\half{{1\over 2}}
\def\hhalf{{1\over 4}}
\def\Tr{{\rm Tr\ }}
\def\im{{\rm Im\hskip0.1em}}
\def\bra#1{{\langle}#1|}
\def\ket#1{|#1\rangle}
\def\vev#1{\langle{#1}\rangle}
\def\CT{\cal T}
\def\Dslash{\rlap{\hskip0.2em/}D}



\begin{titlepage}
\begin{center}
\hfill CERN-TH/97-115\\
\hfill {\tt hep-th/9706032}\\

\vskip 3cm

{ \LARGE \bf Supersymmetric Solutions in Three-Dimensional 
 Heterotic String Theory}

\vskip .3in

{\bf Ioannis Bakas\footnote{\mbox{Permanent Address: Department of Physics,
      University of Patras, GR-26500, Greece}}, Mich\`ele Bourdeau and 
Gabriel Lopes Cardoso
}\footnote{\mbox{Email: \tt 
bakas@nxth04.cern.ch, bourdeau@mail.cern.ch, cardoso@mail.cern.ch}}
\\
\vskip 1cm

{\em Theory Division, CERN, CH-1211 Geneva 23, Switzerland}\\

\vskip .1in

\end{center}

\vskip .2in

\begin{center} {\bf ABSTRACT } \end{center}
\begin{quotation}\noi
We consider the low-energy effective field theory of 
heterotic string theory compactified 
on a seven-torus, and  we construct  electrically charged as well as 
more general solitonic solutions. These solutions preserve 
$1/2,\, 1/4$ and $1/8$ 
of $N=8, D=3$ supersymmetry and have Killing spinors which exist due
to cancellation of holonomies. The associated space--time
line elements do not exhibit the conical structure that often arises in 
$2+1$ dimensional gravity theories.

\end{quotation}
\vskip 5cm
CERN-TH/97-115\\
\hfill June 1997\\
\end{titlepage}
\vfill
\eject

\newpage

\section{Introduction}

The study of three-dimensional gravity theories is interesting in several
respects. For instance, 
general relativity in three space--time dimensions has been a useful 
laboratory
for studying conceptual issues in classical and quantum gravity (see
\cite{carlip1,carlip2} for a review on work on 
$2+1$ dimensional gravity). More recently, the study of 
 duality symmetries of  compactified string theories down to three 
dimensions has provided some information about  
the large internal symmetries  of this sector \cite{sen1,bakas,her}. 
These symmetries are of interest, as they can
yield non-perturbative information about the full string theory.
  
Another interesting aspect that has been recently
 pointed out by Witten \cite{wit1,wit2} is that the vanishing
of the cosmological constant and the absence of a massless dilaton
in four space--time dimensions could be explained by duality between
a supersymmetric string vacuum in three dimensions and a non-supersymmetric
string vacuum in four dimensions.  The observation that in $2+1$ dimensions
the usual connection between supersymmetry of the vacuum and the
bose--fermi degeneracy of the excited states does not hold \cite{wit1,wit2}, 
has been subsequently explored in certain three-dimensional models 
 \cite{bbs,keha}.
Other models that have been studied are supersymmetric
spacetimes in $2+1$ anti-de Sitter supergravity \cite{town1}, 
and some new $2+1$ dimensional Poincar\'e
supergravity theories with central charges and Killing spinors \cite{howe}.
All these considerations add renewed interest
to the study of three-dimensional supergravity theories.

In this paper, we will consider the low-energy effective theory of
heterotic string
theory compactified on a seven-torus \cite{sen1}, and we will construct
various static soliton solutions.  Rather than using the criteria
of the saturation of the Bogomol'nyi bound to characterise these
solutions, we will use the criteria of unbroken supersymmetry
\cite{sen1}.
The construction of these supersymmetric  solutions
will thus be achieved by
solving the associated Killing spinor equations. 
The associated space--time metric 
does not approach flat space--time at infinity, as is the case in four
dimensions, and this renders the existence of covariantly constant spinors
uncertain at first sight, due to the phase acquired by a spinor when 
parallel transported
around a closed curve at infinity. We show, however, that it is possible
to construct such Killing spinors due to the cancellation of the
holonomies. The existence  of non-trivial supercovariantly constant
Killing spinors in asymptotically conical
spacetimes \cite{dejahooft} due to the cancellation of phases
has already been noticed 
in various other three-dimensional
models \cite{town1,bbs,howe,scha,edel,edel1,ggp,keha}.

This paper is organised as follows.  In section 2 we review some properties 
of the low-energy effective action of heterotic string theory
compactified on a seven dimensional torus \cite{sen1}.
In section 3 we present the Killing spinor equations associated to 
the three-dimensional heterotic
low-energy effective Lagrangian.  
Consistency with the Clifford algebra in ten dimensions forces us
to introduce a chirality operator in three dimensions \cite{wetterich} (see
appendix).
In order to be able to do so,
we promote the three-dimensional
Killing spinors to four-component spinors 
(no two-dimensional representation for the three-dimensional
Dirac matrices exists admitting
a gamma matrix anticommuting with all of them).

In section 4 we present static
soliton solutions,
which we obtain by solving the Killing spinor equations along the lines
of \cite{cvetic}.  
We find that the space--time line element differs from the
line element associated with conical geometries \cite{dejahooft}.
We proceed in several steps.  First, we construct electrically charged
solutions.  We take the associated gauge fields to be the ones
arising from the compactification of the heterotic string from ten dimensions
down to three.  
We further restrict the internal metric $G_{mn}$ to be diagonal.
This restriction has the consequence that the electrically charged
solution can, at most, carry
two electric charges associated with
two different $U(1)$ factors.  In subsection 4.1 we construct
electrically charged solutions carrying both charges, and we show that
they
preserve $1/2$ of $N=8, D=3$ supersymmetry.  The associated 
internal metric $G_{mn}$ is constant, whereas the internal
antisymmetric tensor field $B_{mn}$ is zero.  Next, since
the low-energy effective theory is invariant under $O(8,24)$
transformations
of the background fields,
we apply a particular
$O(8,24)$ transformation on the background fields of the electrically 
charged solution, and we obtain two types of 
solitonic solutions which also preserve
$1/2$ of $N=8, D=3$ supersymmetry.  In particular, the type of  solitonic
solutions given  in subsection 4.2.1 has an off-diagonal 
non-constant internal metric $G_{mn}$
as well as a non-vanishing internal antisymmetric tensor field $B_{mn}$.
In addition, the associated gauge field strengths vanish.
Then, we proceed to construct solitonic solutions preserving $1/4$ of 
$N=8, D=3$ supersymmetry, by combining features of the electrically
charged solutions and of the solitonic solutions of subsection
4.2.1.  That is, they have non-vanishing gauge field strengths 
 as well as a non-diagonal non-constant internal metric
and a non-vanishing internal antisymmetric tensor field.  
Finally, this procedure can be generalised to yield
solitonic solutions preserving $1/8$ of 
$N=8, D=3$ supersymmetry.  This is achieved by increasing the number of
non-vanishing entries (blocks) in the $B_{mn}$-field.

In section 5 we repeat the analysis given in section 4, starting
from electrically charged solutions carrying only one electric
charge.  These electrically charged solutions have a
non-constant internal metric $G_{mn}$, as opposed to the ones discussed
in section 4.  We proceed to construct solitonic solutions preserving
$1/2,\; 1/4$ and $1/8$  of $N=8, D=3$ supersymmetry along the line
of section 4.  Here we find in all cases that the internal metric 
$G_{mn}$ is non-constant, but diagonal.

The space--time curvature of each of the solutions
constructed in sections 4 and 5 vanishes at spatial infinity, but the
associated space--time metric does not asymptotically approach either 
a flat metric or an anti-de Sitter metric.  
Thus, these solutions do not describe black hole
solutions in the usual sense \cite{hor1,btz}. 
Our supersymmetric solutions do not appear to interpolate spatially 
between two vacuum-type  supersymmetric  configurations, as  
is the case for the extreme
Reissner--Nordstr\"om metric in four dimensions, for example.
This latter solution interpolates between flat space--time at
spatial infinity and a Bertotti--Robinson metric near the horizon
\cite{gibbons}. 
We  nevertheless refer to our supersymmetric solutions as solitonic
solutions.

In \cite{sen1}, Sen constructed a particular three-dimensional 
 solution  by first considering the fundamental string solution of the four
dimensional theory \cite{dab} and then winding the direction along which 
the string extends
once in the third direction.  In section 6, we construct the associated
Killing spinor in three dimensions, as an application of our formalism.

All solutions discussed in sections 4, 5 and 6 have $H_{\mu \nu \rho} =0$.
In section 7, we  consider solutions to the Killing spinor equations
with $H_{\mu \nu \rho} \neq 0$,
which preserve 1/2 of $N=8, D=3$ supersymmetry.
We  show that all such solutions, with the exception of one, do not
solve the equations of motion.  This should be compared with the 
 common expectation
\cite{boucher,scha,edel} 
that (under some suitable general assumptions) 
every solution to the Killing spinor equations also
solves the equations of motion.

Finally, in section 8, we present our conclusions. 
Our conventions are summarised in the appendix.

\section{The three-dimensional effective action}

\setcounter{equation}{0}

The effective low-energy field theory of the ten-dimensional heterotic string
compactified on a seven-dimensional torus is obtained from reducing the
ten-dimensional $N=1$ supergravity theory coupled to $U(1)^{16}$ super 
Yang-Mills multiplets 
 (at a generic point in the moduli space) \cite{fer,mah,sen1}.
The massless ten-dimensional bosonic fields are the metric $G^{(10)}_{MN}$, 
the anti-symmetric tensor field $B^{(10)}_{MN}$, the $U(1)$ gauge 
fields $A^{(10)I}_M$
and the scalar dilaton $\Phi^{(10)}$ with $(0\leq M,N\leq 9,\quad 
 1\leq I\leq 16)$.
The field strengths are $F^{(10)I}_{MN}=\del_MA^{(10)I}_N-\del_NA^{(10)I}_M$ 
and $H^{(10)}_{MNP}=(\del_MB^{(10)}_{NP}-\half A^{(10)I}_MF^{(10)I}_{NP})+$
cyclic permutations of $M,N,P$.

The bosonic part of the ten dimensional action is

\eqn
{\cal S} &\propto & \int d^{10}x\sqrt{-G^{(10)}}
e^{-\Phi^{(10)}}[{\cal R}^{(10)}
+G^{(10)MN}\del_M\Phi^{(10)}\del_N\Phi^{(10)}\no\\
&& \qquad\qquad\qquad -{1\over 12}H^{(10)}_{MNP}H^{(10)MNP}
-\hhalf F^{(10)I}_{MN}F^{(10)IMN}].
\enq

The reduction to three dimensions \cite{marcus,mah,sen1} introduces 
the graviton
$g_{\mu\nu}$, the dilaton $\phi \equiv \Phi^{(10)} -\ln \sqrt{\det G_{mn}}\,$,
 with $G_{mn}$
the internal 7D metric, 30 $U(1)$ gauge fields 
$A^{(a)}_\mu\equiv (\,A_\mu^{(1)m},\, A^{(2)}_{\mu m},\, A_\mu^{(3)I}\,)
\quad (a=1,\dots,30,\;m=1,\dots ,7,\; I=1,\dots ,16)\;,$
where $\;A_\mu^{(1)m}\;$ are the 7 Kaluza--Klein gauge fields coming from 
the reduction of $G_{MN}^{(10)},\;A_{\mu m}^{(2)}\equiv B_{\mu m} 
+ B_{mn}A_\mu^{(1)n}+\half a_m^IA_\mu^{(3)I}\;$ are the 7 gauge fields coming
from the reduction of $\;B_{MN}^{(10)}\;$ and $\;A_\mu^{(3)I}\equiv A_\mu^I\
-a_m^IA_\mu^{(1)m}\;$ are the 16 gauge fields from  $A_M^{(10)I}$.

 The field strengths $F_{\mu\nu}^{(a)}$ are given by 
$F_{\mu\nu}^{(a)}=\del_\mu A_\nu^{(a)}-\del_\nu A_\mu^{(a)}$. 
Finally, $\;B_{MN}^{(10)}\;$ induces the two form field 
$\;B_{\mu\nu}\;$ with field strength $\;H_{\mu\nu\rho}=\del_\mu 
B_{\nu\rho}-\half A_\mu^{(a)}L_{ab}F_{\nu\rho}^{(b)}$+ cyclic permutations.
 
 The 161 scalars $G_{mn},\,a_m^I$ and $B_{mn}$ can be arranged 
into a $30\times 30$ matrix $M$ (we use here the conventions of \cite{mah})
\eq
M=  \left( \begin{array}{ccc}
G^{-1} & -G^{-1}C & -G^{-1}a^T \\
-C^TG^{-1} & G+C^TG^{-1}C+a^Ta & C^TG^{-1}a^T+a^T\\
-aG^{-1} & aG^{-1}C+a & I_{16}+aG^{-1}a^T
\end{array} \right)\;\;,\label{M} 
\en
where $G\equiv [{G}_{mn}],\,\, C\equiv [\half 
a_m^{I}a_n^{I}+B_{mn}]$ and $a\equiv [a^I_m]$.

We have $MLM^T=L,\quad M^T=M,\quad L^{-1}=L,$ where
\eq
 L =  \left( \begin{array}{ccc}
0 & I_7 & 0\\
I_7 & 0 & 0\\
0 & 0 & I_{16} 
\end{array} \right)\;\;. 
\en

    We use the following ansatz for the Kaluza-Klein 10D vielbein 
$E_M^A$ and inverse vielbein $E_A^M$, in the string frame
\eq
\ E^A_M=  \left( \begin{array}{cl}
e^{\phi}e^\alpha_\mu & A_\mu^{(1)m}e^a_m \\
0 & e^a_m 
\end{array} \right)\;\;,\qquad E_A^M=  \left( \begin{array}{cl}
e^{-\phi}e_\alpha^\mu & -e^{-\phi}e_\alpha^\mu A_\mu^{(1)m} \\
0 & e_a^m 
\end{array} \right)\;\;, 
\en
where $e_m^a$ is the internal  and $e_\mu^\alpha$  
the space--time vielbein in the Einstein frame (the relation between
 string metric $G_{\mu\nu}$ and  Einstein metric $g_{\mu\nu}$
in three dimensions is $G_{\mu\nu}=e^{2\phi}g_{\mu\nu}$).

The three-dimensional action in the Einstein frame is then \cite{mah,sen1}, 
\eqn
{\cal S}&=&\hhalf\int d^3x\sqrt{-g}\bigl\{{\cal R}-g^{\mu\nu}
\del_\mu\phi\del_\nu\phi - 
{1\over 12} e^{-4\phi}g^{\mu\mu'}g^{\nu\nu'}g^{\rho\rho'}H_{\mu\nu\rho}
H_{\mu'\nu'\rho'}\no\\
&& - \hhalf e^{-2\phi}g^{\mu\mu'}g^{\nu\nu'}F_{\mu\nu}^{(a)}
(LML)_{ab}F_{\mu'\nu'}^{(b)}+ {1\over 8}g^{\mu\nu}\Tr (\del_\mu ML\del_\nu
ML)\big\}\;\;,
\enq
where $a=1,\dots,30.$

This action is invariant under the $O(7,23)$ transformations
\eq
M\rightarrow \tilde{\Omega}M\tilde{\Omega}^T,\quad A_\mu^{(a)}\rightarrow
\tilde{\Omega}_{ab}A_\mu^{(b)},\quad g_{\mu\nu}\rightarrow g_{\mu\nu},\quad
B_{\mu\nu}\rightarrow B_{\mu\nu},\quad \phi\rightarrow\phi,\quad
\tilde{\Omega}^TL\tilde{\Omega}=L,\label{07}
\en
where $\tilde{\Omega}$ is a $30\times 30$ $O(7,23)$ matrix.

The equations of motion for $A^{(a)}_\mu,\,\, \phi,$ $H^{\mu\nu\rho}$ 
and $g^{\mu\nu}$ are, respectively,
\eqn
&&\del_\mu(e^{-2\phi}\sqrt{-g}(LML)_{ab}
F^{(b)\mu\nu})+\half e^{-4\phi}\sqrt{-g}\,L_{ab}\,F^{(b)}_{\mu\rho}
H^{\nu\mu\rho}=0\;,\label{ff}\\
&&D_\mu D^\mu\phi +\hhalf e^{-2\phi}F_{\mu\nu}^{(a)}(LML)_{ab}
F^{\mu\nu (b)}+{1\over 6}e^{-4\phi}H^{\mu\nu\rho}H_{\mu\nu\rho}=0\;,
\label{phi}\\
&&\del_\mu(\sqrt{-g}e^{-4\phi}H^{\mu\nu\rho})=0\;,\label{H}\\
&&{\cal R}_{\mu\nu}=\del_\mu\phi\del_\nu\phi + \half e^{-2\phi}
F_{\mu\rho}^{(a)}(LML)_{ab}
F_\nu^{\rho (b)}-{1\over 8}\Tr (\del_\mu ML\del_\nu ML)\label{Ricci}\\
&&\qquad-\hhalf e^{-2\phi}g_{\mu\nu}F_{\rho\tau}^{(a)}(LML)_{ab}
F^{\rho\tau (b)}+\hhalf e^{-4\phi}H_\mu^{\tau\sigma}H_{\nu\tau\sigma}
-{1\over 6}g_{\mu\nu}e^{-4\phi}H^{\tau\sigma\rho}H_{\tau\sigma\rho}
\;.\no
\enq

We note that after dimensional reduction on a seven torus, 
the only massless bosonic fields 
remaining are the spin two (non-propagating) graviton $g_{\mu\nu}$ 
and a set of scalar fields, since in three dimensions vector fields 
are dual to scalar fields. In three dimensions the field $B_{\mu\nu}$ has
no physical degrees of freedom. We will therefore consider backgrounds
where either $H_{\mu\nu\rho}=0$, or $H_{\mu\nu\rho}=
\sqrt{-g}\epsilon_{\mu\nu\rho}\Lambda e^{4\phi}.$  

Let us now consider the case where $H_{\mu\nu\rho}=0$. 
{From} the equations of motion for the gauge
fields $A_\mu^{(a)}$ (\ref{ff}) one can define a  set of scalar 
fields $\Psi^a,\; a=1,\dots,30 ,$ through \cite{sen1}
\eqn
&&\sqrt{-g}e^{-2\phi}g^{\mu\mu'}g^{\nu\nu'}(ML)_{ab}F_{\mu'\nu'}^{(b)}=
 \epsilon^{\mu\nu\rho}\del_\rho\Psi^a,\no\\
&&F^{(a)\mu\nu}={1\over{\sqrt{-g}}}e^{2\phi}(ML)_{ab}
\epsilon^{\mu\nu\rho}\del_\rho\Psi^b\label{psi}\;.\label{sca}
\enq
Then, from the Bianchi identity $\epsilon^{\mu\nu\rho}\del_\mu
F_{\nu\rho}^{(a)}=0$,
\eq
D^\mu(e^{2\phi}(ML)_{ab}\del_\mu\Psi^b)=0.
\en
Following \cite{sen1},
the charge quantum numbers of elementary string
excitations are characterized by a 30 dimensional vector $\vec{\alpha}\in 
\Lambda_{30}$. The asymptotic value of the field strength $F_{\mu\nu}^{(a)}$
associated with such an elementary particle can be calculated to be \cite{sen1}
\eq
\sqrt{-g}F^{(a)tr}\simeq -{1\over{2\pi}}e^{2\phi}M_{ab}\alpha^b.\label{F0r}
\en
The asymptotic form of $\Psi^a$ is then
\eq
\Psi^a\simeq -{\theta\over {2\pi}}L_{ab}\alpha^b + {\rm constant}.
\label{asympsi}
\en
Arranging now the $\Psi$'s into a 30
dimensional column vector, one can define 
 a new $32\times 32$ matrix $\cal{M}$ \cite{sen1} 
\renewcommand{\arraystretch}{0.5}
\eq
 {\cal M}=  \left( \begin{array}{ccc}
M+e^{2\phi}\Psi\Psi^T & -e^{2\phi}\Psi & ML\Psi+\half
e^{2\phi}\Psi(\Psi^TL\Psi)  \\
&& \\
-e^{2\phi}\Psi^T & e^{2\phi} & -\half e^{2\phi}\Psi^TL\Psi\\
&& \\
\Psi^TLM+\half e^{2\phi}\Psi^T(\Psi^TL\Psi) & -\half e^{2\phi}\Psi^TL\Psi
  & e^{-2\phi}+\Psi^TLML\Psi +\hhalf e^{2\phi}(\Psi^TL\Psi)^2
\end{array} \right)\;\;, 
\en
\renewcommand{\arraystretch}{1.0}
where ${\cal M}^T={\cal M},\quad {\cal M}^T{\cal L}{\cal M}={\cal L},\;$
and  $\;{\cal L}\;$ is a $32\times 32$ matrix 
\eq
 {\cal L}=  \left( \begin{array}{ccc}
L & 0 & 0  \\
0 & 0 & 1\\
0 & 1 & 0
\end{array} \right)\;\;. 
\en
Then the action in the Einstein frame can be written as \cite{sen1}
\eq
{\cal S}=\hhalf \int d^3x\sqrt{-g}\left[{\cal R}+{1\over 8}g^{\mu\nu}
\Tr(\del_\mu{\cal  M}{\cal L}\del_\nu{\cal M}{\cal L})\right],
\en

and is invariant under the $O(8,24)$ transformation
\eq
{\cal M}\rightarrow \Omega{\cal M}{\Omega}^T,\qquad
g_{\mu\nu}\rightarrow g_{\mu\nu},
\en
with the  $32\times 32$ matrix $\Omega$ satisfying 
$\Omega^T {\cal L}\Omega ={\cal L}.$
    The low energy effective three dimensional 
field theory becomes then invariant under $O(8,24)$ transformations.

    As explained in \cite{sen1}, this $O(8,24)$ symmetry may be understood  
as a combination of the $O(7,23)$ symmetry (\ref{07}) and the ${\rm SL(2,\RR)}$
symmetry of the four dimensional effective action.
The three dimensional theory may be regarded as arising from
compactification of the  four dimensional theory on a circle, i.e.  consider
the four dimensional theory to be obtained by compactifying the directions
4-9. The three-dimensional theory is then obtained by compactifying the
direction 3 on a circle. Then the ${\rm SL(2,\RR)}$ transformation 
of the four dimensional axion-dilaton complex scalar
field  $\; \lambda\rightarrow (a\lambda +b)/(c\lambda +d)\quad$ 
\cite{sen2} generated by the matrix $ \left( \begin{array}{cc} a & b  
\\c & d\end{array} \right)$ with $ad-bc=1$, 
corresponds to the following transformation on
the three dimensional fields ${\cal M}\rightarrow {\Omega}{\cal  M}{\Omega}^T$
 \cite{sen1} :
\beqa
\Omega =  \left( \begin{array}{ccccccc}
a & 0   & 0 & 0 & 0 & b & 0  \\
0 & I_6 & 0 & 0 & 0 & 0 & 0  \\
0 & 0   & d & 0 & 0 & 0 & -c  \\
0 & 0   & 0 & I_6 & 0 & 0 & 0  \\
0 & 0 & 0 & 0 & I_{16} & 0 & 0  \\
c & 0 & 0 & 0 & 0 & d & 0  \\
0 & 0 & -b & 0 & 0 & 0 & a
\end{array} \right) \;\;\;,\;\;\;\; ad-bc = 1\;\;\;,\label{omega}
\eeqa
with $\Omega$ being a $O(8,24)$ transformation. The full $O(8,24)$ group of
transformations is then generated from the $O(7,23)$ transformations
(\ref{07}) and the ${\rm SL(2,\RR)}$ transformation written above.
In fact, a $O(8,24;\IZ)$ subgroup of this group is 
a symmetry of the full string theory \cite{sen1}.

\section{The Killing spinor equations}

\setcounter{equation}{0}

In ten dimensions, the supersymmetry transformation rules  for 
the gaugini $\chi^I$, dilatino
$\lambda$ and gravitino $\psi_M$ are,  in the string frame, given by 
\cite{julia, stro, cand, liu, peet}
\eqn
&&\delta \chi^I= \half F^I_{MN}\Gamma^{MN}\varepsilon\;\;,\no\\
&&\delta\lambda =-\half \Gamma^M\del_M\Phi\varepsilon
 +{1\over 12} H_{MNP}\Gamma^{MNP}\varepsilon\;\;,\no\\
&&\delta\psi_M=\del_M\varepsilon + {1\over 4}(\omega_{MAB}-\half H_{MAB})
\Gamma^{AB}\varepsilon\;\;.
\enq
These equations become, when reduced to three dimensions in the Einstein
frame,
\eqn
\delta \chi^I&=&\half e^{-2\phi}(F^{(3)I}_{\mu\nu}
+F^{(1)m}_{\mu\nu}a^I_m)
\gamma^{\mu\nu}\varepsilon +  e^{-\phi}\del_\mu a^I_m\gamma^\mu\gamma^4\otimes
\Sigma^m\varepsilon\;\;,\no\\
\delta\lambda &=& -\half e^{-\phi}\del_\mu \{\phi+\ln\det e_m^a\} 
\gamma^\mu\otimes{\bf I}_8\,\varepsilon +{1\over 12}e^{-3\phi}H_{\mu\nu\rho}
\gamma^{\mu\nu\rho}\varepsilon\no\\
&&+\hhalf e^{-2\phi}[-C_{mn}F^{(1)n}_{\mu\nu}+F^{(2)}_{\mu\nu m} -a^I_m
F^{(3)I}_{\mu\nu}]\gamma^{\mu\nu}\gamma^4\otimes\Sigma^m\varepsilon\no\\
&&+\hhalf e^{-\phi}[\del_\mu B_{mn}+\half
(a^I_m\del_\mu a^I_n-a^I_n\del_\mu a^I_m)]\gamma^\mu\otimes\Sigma^{mn}
\varepsilon\;\;,\no\\
\delta{\psi}_\mu &=&\del_\mu\varepsilon + {1\over 4}\omega_{\mu\alpha\beta}
\gamma^{\alpha\beta}\varepsilon 
 +\hhalf (e_{\mu\alpha}e_\beta^\nu \!-\!e_{\mu\beta}e_\alpha^\nu)
\del_\nu\phi\gamma^{\alpha\beta}\varepsilon + {1\over 8}(e^n_a\del_\mu 
e_{nb}\!-\!e^n_b\del_\mu e_{na}){\bf I}_4\otimes\Sigma^{ab}\varepsilon \no\\
&&\!\! -{1\over 8}e^{-2\phi}H_{\mu\nu\delta}\gamma^{\nu\delta}\varepsilon
\!-\!{1\over 4}e^{-\phi}[e^m_a F^{(2)}_{\mu\nu(m)} 
- e_{ma} F^{(1)m}_{\mu\nu}]\gamma^\nu\gamma^4\otimes\Sigma^a\varepsilon
\!-\!{1\over 8}[\del_\mu B_{mn}\! +\!\half (a^I_m\del_\mu a^I_n \no\\
&&-a^I_n\del_\mu a^I_m)\;]\;{\bf I}_4
\otimes\Sigma^{mn}\varepsilon
-\hhalf e^{-\phi}[-C_{mn}F_{\mu\nu}^{(1)n}-a^I_m
F_{\mu\nu}^{(3)I}\;]\;\gamma^\nu\gamma^4\otimes
\Sigma^m\varepsilon\;\;,\no\\
\delta\psi_d &=&-{1\over 4}e^{-\phi}(e_d^m\del_\mu e_{ma}\!+\!e_a^m\del_\mu 
e_{md})\gamma^\mu\gamma^4\otimes\Sigma^a\varepsilon
-{1\over 8}e^{-2\phi}e^m_d[-C_{mn}F^{(1)n}_{\mu\nu}
\!-\!a^I_mF^{(3)I}_{\mu\nu}]\gamma^{\mu\nu}\varepsilon\no\\
&&+{1\over 4}e^{-\phi}e^m_de^n_a(\del_\mu B_{mn}+\half 
(a^I_m\del_\mu a^I_n - a^I_n\del_\mu a^I_m))\gamma^\mu
\gamma^4\otimes\Sigma^a\varepsilon\no\\
&&-{1\over 8}e^{-2\phi}[\;e_{md}F_{\mu\nu}^{(1)m}
+e_d^mF^{(2)}_{\mu\nu m}\;]\gamma^{\mu\nu}\varepsilon\;\;,\label{killing}
\enq
where $\delta\psi_d\equiv e_d^m\delta\psi_m$ denotes the variation of the
internal gravitini, and where we have suppressed
the label $i$  indicating the  supersymmetries $(i=1,\dots,8)$
 as well as the index $A$ for the space--time dimensionality of the spinors
 ($A=1,\dots, 4)$ (see appendix).

We would now like  to construct static solutions to the Killing spinor 
equations by taking the supersymmetry variations of the fermionic fields 
to zero. This will insure that the bosonic configuration so obtained 
will be supersymmetric.

We will take the space--time metric to be diagonal. In all cases, with the
exception of the one discussed in section \ref{cosmic1}, the space--time
metric will be given by 

\eq
ds^2=-V(r)dt^2 +V(r)^{-1}dr^2 +R^2(r)d\theta^2\;\;,\label{gg}
\en
for which
\eqn
&&\omega_{t\alpha\beta}\,\gamma^{\alpha\beta}
=-2\sqrt{V}\,\del_r\sqrt{V}\,\gamma^{01}, 
\quad \omega_{r\alpha\beta}\gamma^{\alpha\beta}=0,\quad
\omega_{\theta \alpha\beta}\,\gamma^{\alpha\beta}=-2\sqrt{V}\,\del_r R\,
\gamma^{12}\;,\no\\
&& (e_{\mu\alpha}\,e_\beta^r -e_{\mu\beta}\,e_\alpha^r)
\del_r\phi\gamma^{\alpha\beta}\;\quad\; =\quad-2V\,\del_r\phi \,\gamma^{01},
\quad\quad{\rm for }\quad\mu=t,\no\\
&&\qquad\qquad\qquad \qquad\qquad\qquad = \quad 0,
\quad\qquad\qquad\qquad {\rm for}\quad \mu=r,\no\\
&&\qquad\qquad\qquad\qquad\qquad \qquad =\quad - 2R\sqrt{V}\,
\del_r\phi\,\gamma^{12}\quad{\rm for}\quad \mu=\theta\;.\label{con}
\enq

Then, the Ricci tensor has the following non-zero components
\eqn
{\cal R}_{tt}&=& {V \over 2} (V''+ V' {{R'}\over R})\;,\no\\
{\cal R}_{rr}&=&-{V''\over{2V}} -{V'R'\over {2V R}} 
 - {{R''}\over R}\;,\no\\
{\cal R}_{\theta\theta}&=&-V R R''-V'R' R\;\;,\label{ric}
\enq
and the curvature scalar is given by
\eq
{\cal R}=g^{\mu\nu}{\cal R}_{\mu\nu}=-V''-{{2V'R'}\over R}-{{2VR''}\over
  R}\;,
\en
where $V'=\del_r V,\;\;R'=\del_r R.$

In all cases, we will make the following ansatz for the Killing spinors
\eq
\varepsilon=\epsilon\otimes\chi\label{spinor}\;\;,
\en
where $\epsilon^T=(\epsilon_1,\,\epsilon_2,\,\epsilon_3,\,\epsilon_4)$ is a
SO(1,2) spinor and $\chi$ is a SO(7) spinor of the internal space. 
In all cases, with the exception of the ones discussed in sections
\ref{cosmic1} and \ref{hmu}, we will be able to solve the 
Killing spinor equations 
by imposing the following two conditions on $\epsilon$:
\eqn
\gamma^1\epsilon &=&ip\;\gamma^2\;\epsilon\;\;,\label{cond1}\\
\gamma^1\epsilon &=&\tilde{p}\;J\gamma^2\gamma^4\;\epsilon\;\;,\label{cond2}
\enq
where $p=\pm\;,\;\;\tilde{p}=\pm\;.$

It follows that 
\beqa
\epsilon = \tilde{\epsilon}(r,\theta) \,
\left( \begin{array}{c}
 i p \\
1 \\
 {\tilde p}\\
-i p {\tilde p}
\end{array} \right) \;\;\;,\label{spinor1}
\eeqa
and, hence, $\epsilon$ contains only two real independent degrees of
freedom. $\chi$, on the other hand, contains eight real degrees of freedom;
thus there are a priori a total of 16 real degrees of freedom. These will
be further reduced by conditions on $\chi$ specific to each case
considered. Up to three such independent 
conditions $(m=1,2,3)$ can be imposed on
$\chi$, thereby allowing for the construction of solutions preserving
$1/{2^m}$ of the $N=8, D=3$ supersymmetry.

In all cases where $H_{\mu\nu\rho}=0$, we find that
\eq
\tilde{\epsilon}=e^{{\phi}\over 2}\;e^{iY(r,\;\theta)}\;\;,\label{spinor2}
\en
up to a multiplicative constant.

\section{Supersymmetric solutions with $\vec{\alpha}^2\neq 0$ }

\setcounter{equation}{0}

In this section, we will consider a particular class of solutions to the
Killing spinor equations, namely solutions for which
$\vec{\alpha}^2=\alpha^TL\alpha \neq 0$. We will construct solutions which
preserve $1/2^m$ of $N=8, D=3$ supersymmetry, where $m=1,2,3$.
The solutions are obtained with $H_{\mu\nu\rho}=0$ and $a_m^I=0$. 

We will find that the space--time metric (\ref{gg}) is given in terms of
\beqa
V=1,\qquad\qquad R = a\; r^{1 - \frac{\gamma}{2}} \;\;,
\eeqa
and that the dilaton is given by
\beqa
e^{2 \phi} = r^{-\gamma} \;\;\;,
\eeqa
where
\beqa
\gamma = \frac{2}{n+1} \;\;\;\;,\;\;\;\; 
a = \frac{\sqrt{|\alpha_i \, \alpha_{i+7}|}}{\gamma \, \pi}    \;\;\;.
\eeqa
By the coordinate transformation $r=({\gamma\over 2})^{2\over\gamma}\;(a\ln
  \tilde{r})^{2\over\gamma}\, , \,
1 \leq {\tilde r} \leq \infty$, 
the associated space--time metric can be put
  into the form
\eq
ds^2=-dt^2+ a^{4\over\gamma}\;({\gamma\over
    2})^{{2(2-\gamma)}\over\gamma}\;
{{(\ln\tilde{r})}\over{\tilde{r}^2}}^{{2(2-\gamma)}\over\gamma}
(d\tilde{r}^2+\tilde{r}^2d\theta^2)\label{met}.
\en
This differs from the line element associated with conical geometries
\cite{dejahooft}.

The curvature scalar, ${\cal R}=g^{\mu\nu}{\cal R}_{\mu\nu}$, 
is computed to be
\eq
{\cal R}\,=\,\gamma(1-{\gamma\over 2})\;{1\over
  {r^2}}\,=\,{{2n}\over{(n+1)^2}}\;{1\over {r^2}}\;\;. 
\en

\subsection {Electrically charged solutions\label{2c}}

We will first consider  the case where the internal vielbein 
$e_m^a$ is diagonal and given by  $e_m^a=\delta^m_a e_m(r)$. 
We will also take   $\phi=\phi (r)$ and $B_{mn}=0$.

The Killing  equations (\ref{killing}) reduce to 
\eqn
\delta \chi^I&=&\half e^{-2\phi}F^{(3)I}_{\mu\nu}
\gamma^{\mu\nu}\varepsilon\;\;,\no\\
\delta\lambda &=&-\half e^{-\phi}\del_\mu \{\phi + \ln \det e_m^a \}
\gamma^\mu\varepsilon
+\hhalf e^{-2\phi}e^m_d F^{(2)}_{\mu\nu m}\gamma^{\mu\nu}\gamma^4
\otimes\Sigma^d\varepsilon\;\;,\no\\
\delta{\psi}_\mu &=&\del_\mu\varepsilon + {1\over 4}\omega_{\mu\alpha\beta}
\gamma^{\alpha\beta}\varepsilon +\hhalf (e_{\mu\alpha}e_\beta^r 
-e_{\mu\beta}e_\alpha^r) \del_r\phi\gamma^{\alpha\beta}\varepsilon\no\\ 
&& -{1\over 4}e^{-\phi}[e^m_a F^{(2)}_{\mu\nu(m)} 
- e_{ma} F^{(1)m}_{\mu\nu}]\gamma^\nu\gamma^4\otimes\Sigma^a\varepsilon\;\;,\\
\delta\psi_d &=&-{1\over 2}e^{-\phi}e^m_d\del_\mu e_{md}
\gamma^\mu\gamma^4\otimes\Sigma^d\varepsilon
-{1\over 8}e^{-2\phi}[e^m_d F^{(2)}_{\mu\nu m} +e_{md} F^{(1)m}_{\mu\nu}]
\gamma^{\mu\nu}\varepsilon\;\;.\no
\enq
Using (\ref{gg}), as well as (\ref{con}), we get
\eqn
\delta \chi^I&=&\half e^{-2\phi}F^{(3)I}_{\mu\nu}
\gamma^{\mu\nu}\varepsilon\;\;,\no\\
\delta \lambda &=&-\half e^{-\phi} \del_r\{\phi + \ln \det e_m^a\}
\gamma^r\,\varepsilon +\half e^{-2\phi}e^m_a F_{trm}^{(2)}
\gamma^{tr}\gamma^4\otimes\Sigma^a\varepsilon\;\;,\no\\
\delta{\psi}_t&=&-\half [\sqrt{V}\del_r\sqrt{V}+V
\del_r\phi]\gamma^{tr}\varepsilon -\hhalf 
e^{-\phi}[e^m_aF_{trm}^{(2)}-e_{ma}F_{tr}^{(1)m}]\gamma^r\gamma^4
\otimes\Sigma^a\varepsilon\;\;,\no\\
\delta {\psi}_r&=&\del_r\varepsilon 
-\hhalf e^{-\phi}[e^m_a F_{rtm}^{(2)}-e_{ma}F_{rt}^{(1)m}]\gamma^t\gamma^4
\otimes\Sigma^a\varepsilon\;\;,\\
\delta{\psi}_\theta &=&\del_{\theta}\varepsilon -\half [\sqrt{V}
\del_r R +R\sqrt{V}\del_r\phi]
\gamma^{12}\varepsilon\;\;,\no\\
\delta \psi_d&=& -\hhalf e^{-2\phi}[e^m_dF_{trm}^{(2)} + e_{md}
F_{tr}^{(1)m}]\gamma^{tr}\varepsilon -\half e^{-\phi}\del_r\ln e_m^d
\gamma^r\gamma^4\otimes\Sigma^{d}\varepsilon,\qquad d=1,\dots 7.\no
\enq
In the following, we will set $F_{\mu\nu}^{(3)I}=0,\quad I=1,\dots,16.$

We take the Killing spinor $\varepsilon$ to be given by (\ref{spinor}).

Let us now determine how many electric charges can be non-zero.
Let us assume  that $\Sigma^a\chi = \eta\,\chi$. 
Since $(\Sigma^a)^2=1,\quad \eta=\pm 1$. Suppose now that we also have
 $\Sigma^b\chi=\eta\,\chi$ with $a\neq b$. 
Then, $\Sigma^a\Sigma^b\,\chi
=\,\chi$. Since however $(\Sigma^a\Sigma^b)^2=-1\;$ (for $a\neq b)$,
we must have $\Sigma^a\Sigma^b\,\chi =\pm i M\,\chi$, 
where $M^2={\bf I}$. Therefore the above assumption $a\neq b$ 
is not valid. So, out of the 14 remaining  electric charges,
only two are non-zero, one of them arising from the Kaluza-Klein sector and
the other from the two-form gauge fields. 

For concreteness, we choose $a=2$, and hence, the two non-vanishing charges
are $\alpha_2$ and $\alpha_9$ (see equation \ref{F0r}). 
Note that $\vec{\alpha}^2\neq 0$.

We now set $\Sigma^2\chi = \,\chi$. Then
\eqn
\delta \lambda &=&-\half e^{-\phi} \del_r\{\phi +\ln\det e_m^a\}\sqrt{V}
\gamma^1\varepsilon -\half e^{-2\phi}\sqrt{G^{22}}\, F_{tr2}^{(2)}\,J\,\gamma^2
\gamma^4\varepsilon\;\;,\no\\
\delta{\psi}_t &=& \half [\sqrt{V}\del_r\sqrt{V}+V
\del_r\phi]\,J\,\gamma^2\varepsilon
-\hhalf e^{-\phi}\sqrt{V}[\sqrt{G^{22}}\,F_{tr2}^{(2)}-\sqrt{G_{22}}\,
F_{tr}^{(1)2}]\,
\gamma^1\gamma^4\varepsilon\;\;,\no\\
\delta {\psi}_r&=&\del_r\varepsilon + \hhalf e^{-\phi}\sqrt{V^{-1}}
[\sqrt{G^{22}}\, F_{tr2}^{(2)}-\sqrt{G_{22}}\,F_{tr}^{(1)2}]\,
\gamma^0\gamma^4\varepsilon\;\;,\no\\
\delta{\psi}_{\theta} &=&\del_{\theta}\varepsilon -\half [\sqrt{V}
\del_r R +R\sqrt{V}\del_r\phi]\,J\,\gamma^0\varepsilon\;\;,\\
\delta \psi_2 &=& \hhalf e^{-2\phi}[\sqrt{G^{22}}\,F_{tr2}^{(2)}
+\sqrt{G_{22}}\,F_{tr}^{(1)2}]\,J\gamma^2\varepsilon - \half e^{-\phi}\sqrt{V}
\del_r\ln \sqrt{G_{22}}\,\gamma^1\gamma^4\varepsilon\, ,\no \\
\delta\psi_d &=& - \half e^{-\phi}\sqrt{V}
\del_r\ln \sqrt{G_{dd}}\,\gamma^1\gamma^4\varepsilon\;\; .\no
\enq
In order for the equations to be compatible, we will impose conditions 
(\ref{cond1}) and (\ref{cond2}) on the four-dimensional spinor $\varepsilon$. 

Setting the variations of the supersymmetry equations to zero, 
we have 
\eqn
\del_r\{\phi +\ln \det  e_m^a\}\sqrt{V}
\;&=&\;-\tilde{p}\, e^{-\phi} \,\sqrt{G^{22}}\, F_{tr2}^{(2)}\;\;,
\label{lamb}\\
 \sqrt{V}\,[\half\del_r\ln V+\del_r\phi]\;&=&\;
 {{\tilde{p}}\over 2}\, e^{-\phi}\,[\sqrt{G_{22}}\,F_{tr}^{(1)2}-
\sqrt{G^{22}}\,F_{tr2}^{(2)}]\;\;,\label{psit}\\
 \sqrt{V}\,\del_r\ln \sqrt{G_{22}}\;&=&\;-{{\tilde{p}}\over 2}\,e^{-\phi} 
[\sqrt{G^{22}}\,F_{tr2}^{(2)}+\sqrt{G_{22}}\,F_{tr}^{(1)2}]\;\;,\label{psid}\\
\del_r\,\tilde{\epsilon} \;&=&\;{{\tilde{p}}\over 4}\,
{{e^{-\phi}}\over{\sqrt{V}}}
 \,[\sqrt{G_{22}}\,F_{tr}^{(1)2}-\sqrt{G^{22}}F_{tr2}^{(2)}]\,
\tilde{\epsilon}\;\;,\label{psir}\\
\del_{\theta}\,\tilde{\epsilon} \;&=&\; -{{ip}\over 2}\, [\sqrt{V}
\del_r R +R\sqrt{V}\del_r\phi]\,\tilde{\epsilon}\;\;,\label{psitheta}\\
 G_{dd} &=& {\rm constant}, \qquad d\neq 2\;\;.\label{int}
\enq
{From} (\ref{psit}), (\ref{psir}) and (\ref{psitheta}), one has
\eqn\label{delr}
&&\del_r \,\tilde{\epsilon}-\half\, [\half\del_r\ln V +
\del_r\phi ]\,\tilde{\epsilon}=0\;\;,\label{psir1}\\
&&\del_\theta \,\tilde{\epsilon}+ip\,
\half\, [\sqrt{V}\del_r R +R\sqrt{V}
\del_r\phi ]\,\tilde{\epsilon}=0\;\;.\label{psitheta1}
\enq
In order for these equations to be compatible with respect to 
 the mixed derivative
$\del^{2}_{\,r\theta}$,  we need to impose
$ {{\del }\over{\del r}}[\sqrt{V}\del_r R +R\sqrt{V}
\del_r\phi ]=0$. In the following, we will set $[\sqrt{V}\del_r R +R\sqrt{V}
\del_r\phi ]=0$, and hence $\del_\theta\tilde{\varepsilon}=0$. Then it
 follows that 
\eq
\del_r\phi=-{{R'}\over R}\qquad\longrightarrow\qquad R=ae^{-\phi}.
 \en
We note here, however, that if the spinor were to have a  phase
of the form $e^{i\eta\theta}$, $\eta$ would have to be ${(2n+1)/2}$ (with
$n$ integer), such that $e^{i\eta(\theta=2\pi)}=-e^{i\eta (\theta=0)}$ 
\cite{hen1,town1}, so we would need to have 
$[\sqrt{V}\del_r R +R\sqrt{V}\del_r\phi ]=-(2n+1)p$.
 
We take the two electric fields to be given by
\eqn
F^{(1)m}_{tr}&=&{1\over {2\pi}}{{e^{2\phi}\over R}}\,G^{22}\,
 \alpha_2\;\;,\qquad\qquad m=2\;\;,\no\\
F^{(2)}_{trm}&=&{1\over {2\pi}}{{e^{2\phi}\over R}}\,G_{22}\,
 \alpha_9 \;\;,
\enq
which is consistent with the asymptotic behavior given in (\ref{F0r}).

We now look for solutions with the internal metric $G_{22}$ constant, 
noting here that a more
 general internal metric could be generated by $O(7,23)$ rotations from
 this one.
Then  equation (\ref{psid}) can be solved by
\eq
 \sqrt{G^{22}}\;F^{(2)}_{tr2}\,=\,-\sqrt{G_{22}}\;F^{(1)2}_{tr}
\qquad\qquad \longrightarrow \qquad\qquad
G_{22}=-{{\alpha_2}\over {\alpha_9}}\;\;.
\en
It follows that $\alpha_2$ and $\alpha_9$ have opposite signs. We will in
the following denote the signs of $\;\alpha_2,\;\alpha_9\;$ by
$\;\eta_{\alpha_2},\;\eta_{\alpha_9}$.

It further follows by inspection of (\ref{psit}), (\ref{psid}) 
and  (\ref{lamb})  that $\del_r\sqrt{V}=0$, 
thus $V$ is a constant which  we set equal to 1.

One can now solve straightforwardly for $\phi$ from
(\ref{lamb}) and (\ref{psit}).
By doing so, we find that   $\tilde{p}=-\eta_{\alpha_2}$ as well as
\eq
e^{2\phi}=
{{c^2}\over{|r-r_0|}},\qquad R={a\over c}\sqrt{|r-r_0|},\qquad 
{a\over {c^2}}={1\over \pi}\sqrt{|\alpha_2\alpha_9|}\;\;,\label{sol} 
\en
where $r_0$ and $c$ are integration constants which  will be set to zero
and one, respectively, from now on.

Note that the  coupling constant ${g^2}=e^{2\phi}\longrightarrow 0\,$ 
as $\, r\rightarrow\infty$.

The space--time metric is then of the form
\eq
ds^2=-dt^2+dr^2+a^2\,r\, d\theta^2\label{metric}
\en
or, equivalently, 
\eq
ds^2= - dt^2 + {{a^4}\over 4}\left({{\ln \tilde{r}}
\over {\tilde {r}}}\right)^2(d\tilde{r}^2 + \tilde{r}^2d\theta^2)\,,
\en
where $r={{a^2}\over 4}(\ln {\tilde r})^2.$

The behavior of the spinor $\tilde{\epsilon}$ can also be determined.
We have from  (\ref{psir1}) and  (\ref{psitheta1})
\eqn
&&\del_r \,\,\tilde{\epsilon}
- \half \del_r \phi \,\,\tilde{\epsilon}=0\;\;,\no\\
&&\del_\theta \,\,\tilde{\epsilon}=0\;\;,
\enq
which can easily be solved by
\eq
\tilde{\epsilon}\,=\,e^{{\phi}\over 2}\, \,=\,
r^{-1/4}\;\;,\label{elspinor}
\en
up to a multiplicative constant.
The existence of such a Killing spinor is made possible due  to
the cancellation of holonomies, that is due to a cancellation between the
spin connection and a term involving the dilaton (see equation
(\ref{psitheta1})).  

This solution preserves 1/2 of the $N=8, D=3$ supersymmetry.

Computing the curvature ${\cal R}=g^{\mu\nu}{\cal R}_{\mu\nu}$, we have
${\cal R}={1\over{2r^2}}$ which blows up at $r=0$ but  goes to
zero at $r\rightarrow \infty$. 

Let us  now consider the equations of motion (\ref{Ricci}):
\eqn
{\cal R}_{tt}&=& \half e^{-2\phi}F_{tr}^{(a)}F_t^{r(a)}-\hhalf e^{-2\phi}
       g_{tt}F_{\alpha\beta}^{(a)}F^{\alpha\beta (a)}=0\;\;,\no\\
{\cal R}_{rr}&=&\del_r\phi\del_r\phi + \half e^{-2\phi}F_{rt}^{(a)}(LML)_{aa}
     F_r^{t(a)}-\hhalf e^{-2\phi}g_{11}F_{\alpha\beta}^{(a)}(LML)_{aa}
F^{\alpha\beta (a)}\no\\
&-&{1\over 8}\Tr(\del_rML\del_rML)\,=\,(\del_r\phi)^2\;\;,\no\\
{\cal R}_{\theta\theta}&=&-\half e^{-2\phi}R^2F_{tr}^{(a)}(LML)_{aa}
      F^{tr(a)}\no\\
      &=&{1\over 8\pi^2}e^{2\phi}[\sqrt{G^{22}}\;(\alpha_2)^2
                 +\sqrt{G_{22}}\;(\alpha_9)^2]\;\;.
\enq
Using now (\ref{ric}), it can be checked that our solution (\ref{sol})
solves the equations of motion.

\subsection{Soliton solutions preserving $N=4$ supersymmetry\label{n4}}

\subsubsection{Case $\alpha_2 \neq 0, \alpha_ 9  \neq 0$\label{n41}}

Here, we will discuss the soliton solution which is obtained by dualizing 
the charged solution discussed in subsection \ref{2c}.  
That is, we will utilize the $O(8,24)$ transformation $\Omega$ given in
(\ref{omega}) to generate the dual background 
${\cal M} \rightarrow {\tilde {\cal M}} = \Omega {\cal M} \Omega^T $.
We will, for simplicity, set the transformation parameter 
$d$ to 
$d=0$ in the following, so that $bc=-1$.

Recall that 
the bosonic background fields 
of the charged solution discussed in subsection \ref{2c}
are given by
$\phi,\, (G_{mn}) = {\rm diagonal} 
\left( {G}_{11}, G_{22}, \dots, {G}_{77}\right), \, 
G_{22} = |\alpha_2/\alpha_9|,
B_{mn} = 0, a_m^I = 0$
and $\Psi^T = ( 0 , \Psi_2, 0 , \dots, 0, \Psi_9, 0, \dots, 0)\,=\,
 ( 0 ,\, -{\theta\over{2\pi}}\,\alpha_9\, , 0 , \dots, 0,
 \,-{\theta\over{2\pi}}\,\alpha_2 \,, 0, \dots, 0).$
The dual background fields 
are then given by
\beqa
{\tilde G}^{-1} &=&  
\left( \begin{array}{ccccc}
{\tilde G}^{11} & {\tilde G}^{12}& 0 & \cdots & 0\\
{\tilde G}^{21} & {\tilde G}^{22}& 0 & \cdots & 0 \\
0 & 0 & {\tilde G}^{33} & 0 \cdots & \vdots \\
\vdots & & & \ddots & \\
0& & \cdots  & 0 & {\tilde G}^{77}
\end{array} \right)= 
\left( \begin{array}{ccccc}
b^2 e^{2 \phi} & - b e^{2 \phi} \Psi_2  & 0 & \cdots & 0\\
-b e^{2 \phi} \Psi_2 & G^{22} +  e^{2 \phi} \Psi_2^2 &
0 & \cdots & 0 \\
0 & 0 & {G}^{33} & 0 \cdots & \vdots \\
\vdots & & & \ddots & \\
0& & \cdots  & 0 & {G}^{77}
\end{array} \right) \;, \nonumber\\
{\tilde B} &=&  ({\tilde B}_{mn}) = 
\left( \begin{array}{ccccc}
0 &{\tilde B}_{12} & 0 & \cdots &0\\
{\tilde B}_{21} & 0 \\
0& & 0 & & \vdots\\
\vdots& & & \ddots \\
0& &\cdots & & 0
\end{array} \right)= 
\left( \begin{array}{ccccc}
0 & - c \Psi_9 & 0 & \cdots &0\\
c \Psi_9 & 0 \\
0& & 0 & & \vdots\\
\vdots& & & \ddots \\
0& &\cdots & & 0
\end{array} \right) 
\label{gbdual}
\eeqa
as well as 
\beqa
e^{2 {\tilde \phi}} &=& c^2 G^{11} \;\;\;\;\;\;,\;\;\;\;\;\;
{\tilde \Psi}=
\left( \begin{array}{c}
{\tilde \Psi}_1 \\
{\tilde \Psi}_2 \\
\vdots \\
{\tilde \Psi}_9 \\
\vdots
\end{array} \right) = 
\left( \begin{array}{c}
-a/c\\
0 \\
\vdots \\
0 \\
\vdots
\end{array} \right) \;\;\;\;\;\;,\;\;\;\;\;\; {\tilde a}_m^I = 0 \;\;\;\;.
\eeqa
Note that the associated gauge field strengths $F_{\mu \nu}^{(a)}$
are all zero for this solitonic solution.  The internal inverse vielbein
${\tilde e}_a^{m}$ associated to (\ref{gbdual}) 
is given by
\beqa
{\tilde e}_a^{m}&=&  
\left( \begin{array}{ccccc}
 b e^{\phi} & - e^{\phi} \Psi_2
& 0 & \cdots & 0\\
0 & \sqrt{{G}^{22}}& 0 & \cdots & 0 \\
0 & 0 & \sqrt{{G}^{33}} & 0 \cdots & \vdots \\
\vdots & & & \ddots & \\
0& & \cdots  & 0 & \sqrt{{G}^{77}}
\end{array} \right) \;\;\;.
\eeqa

Note that the space--time metric is duality invariant and hence given as
before (see (\ref{metric})).

Next, we would like to determine the Killing spinor 
associated with the soliton background (\ref{gbdual}).
The Killing spinor equations (\ref{killing}) now take the form
\beqa
\delta \chi^I&=& 0 \;\;\;,\;\;\; \nonumber\\
\delta\lambda &=& -\half e^{-{\tilde \phi}}\del_\mu
\log \det {\tilde e}_m^a 
\gamma^\mu \varepsilon 
+\hhalf e^{-{\tilde \phi}}\del_\mu {\tilde B}_{mn}
\gamma^\mu\otimes\Sigma^{mn}
\varepsilon \;\;\;,\;\;\;   \label{duallam}\\
\delta \psi_\mu &=&\del_\mu\varepsilon + {1\over 4}\omega_{\mu\alpha\beta}
\gamma^{\alpha\beta}\varepsilon 
+ {1\over 8}({\tilde e}^n_a\del_\mu 
{\tilde e}_{nb}-{\tilde e}^n_b\del_\mu {\tilde e}_{na})
\Sigma^{ab}\varepsilon 
- {1\over 8}\del_\mu {\tilde B}_{mn}
\Sigma^{mn}\varepsilon \;\;\;,  \label{dualgrav}\\
\delta\psi_d &=&-{1\over 4}e^{-{\tilde \phi}}
({\tilde e}_d^m\del_\mu {\tilde e}_{ma}\!
+\!{\tilde e}_a^m\del_\mu 
{\tilde e}_{md})\gamma^\mu\gamma^4\otimes\Sigma^a\varepsilon 
\;+\; {1\over 4} e^{-{\tilde \phi}}
{\tilde e}^m_d {\tilde e}^n_a \del_\mu {\tilde B}_{mn}\!
\gamma^\mu\gamma^4\!\otimes\!\Sigma^a\varepsilon .\nonumber\\
\label{dualintgrav}
\eeqa
The Killing spinor $\varepsilon = \epsilon \otimes \chi$ will be
taken to satisfy (\ref{cond1}) and (\ref{cond2}).
For the solitonic background under consideration, the 
vanishing of the Killing spinor
equation (\ref{duallam}) then yields
\beqa
- \partial_r \log \det {\tilde e}_m^a \gamma^1 \varepsilon
+b \frac{\sqrt{G^{22}}}{R} e^{\phi}\partial_{\theta} {\tilde B}_{12} \gamma^2 
\Sigma^{12} \varepsilon = 0 \;\;\;,
\label{dualkl2}
\eeqa
which can be solved by demanding that the Killing spinor 
$\varepsilon =  \epsilon \otimes \chi$ should also satisfy
\beqa
\Sigma^{12} \, \chi = q \, i \, \chi \;\;\;,\;\;\;
q = \pm  \;\;\;.
\label{dualchi}
\eeqa
Then, equation (\ref{dualkl2}) turns into
\beqa
- p \,\partial_r \log \det {\tilde e}_m^a 
+b\, q\, 
\frac{\sqrt{G^{22}}}{R} e^{\phi}\partial_{\theta} {\tilde B}_{12} 
= 0 \;\;\;,
\eeqa
which is indeed satisfied, provided one takes $q = - p \, 
\eta_{\alpha_2}$, where $\eta_{\alpha_2}$ denotes the sign of the charge
$\alpha_2$, $\eta_{\alpha_2} = {\rm sign} \, \alpha_2$.

Next, consider solving the Killing spinor equations (\ref{dualgrav}).
We will again make the ansatz that the Killing spinor is static, that is
$\varepsilon = \varepsilon (r, \theta)$.  Then, the equation
$\delta \psi_t = 0$ is automatically satisfied.  
The condition $\delta \psi_r =0$, on the other hand,  yields 
\beqa
\partial_r \varepsilon = 0 \;\;\;.
\eeqa
Finally, the condition $\delta \psi_{\theta} =0$
results in 
\beqa
\partial_{\theta} \varepsilon - \frac{1}{2} \partial_r R \gamma^{12} 
\varepsilon + \frac{1}{4} \sqrt{G_{22}} e^{\phi} \partial_{\theta} \Psi_2
\Sigma^{12} \varepsilon - \frac{1}{4} \sqrt{G^{22}} e^{\phi}
\partial_{\theta} \Psi_9 \Sigma^{12} \varepsilon =0 \;\;\;.
\label{thgr}
\eeqa
Inserting the conditions (\ref{cond1}) and (\ref{dualchi}) into
(\ref{thgr}) yields that
\beqa
\partial_{\theta} \varepsilon =0 \;\;\;.
\eeqa
Thus, it follows that the Killing spinor $\varepsilon$ is constant.

Finally, it can be checked that the Killing spinor equations 
(\ref{dualintgrav}) for $\delta \psi_1 $ and $\delta \psi_2$
are also satisfied.

The solitonic background under consideration preserves $1/2$ of $N=8$
supersymmetry.

\subsubsection{Case $\alpha_1 \neq   0, \alpha_8 \neq 0$}

Next, we will discuss a different soliton solution, which will be obtained
by dualizing 
a charged solution with 
bosonic background fields 
$\phi, \; (G_{mn}) = {\rm diagonal} 
\left( {G}_{11},  \dots, {G}_{77}\right), \;
G_{11} = |\alpha_1/\alpha_8|, \;
B_{mn} = 0, \;a_m^I = 0$
and $\Psi^T = ( \Psi_1, 0 , \dots, 0, \Psi_8, 0, \dots, 0)
\,=\,
 ( -{\theta\over{2\pi}}\,\alpha_8\, , 0 , \dots, 0,
 \,-{\theta\over{2\pi}}\,\alpha_1 \,, 0, \dots, 0).$
This charged solution is similar to the one discussed in subsection \ref{2c}.

The dual background fields can be read off from
${\tilde {\cal M}} = \Omega {\cal M} \Omega^T $, where $\Omega$ is 
again given by (\ref{omega}).
We will, for simplicity, set the transformation parameter 
$d$ to 
$d=0$ in the following, so that $bc=-1$. 
For this choice, the
dual background fields 
are given by
${\tilde B} =  ({\tilde B}_{mn}) = 0, \;
{\tilde a}_m^I = 0$,
\beqa
{\tilde G} &=&  
\left( \begin{array}{ccccc}
{\tilde G}_{11} & 0 & 0  & \cdots & 0\\
0 & {\tilde G}_{22}& 0 & \cdots & 0 \\
0 & 0 & {\tilde G}_{33} & 0 \cdots & \vdots \\
\vdots & & & \ddots & \\
0& & \cdots  & 0 & {\tilde G}_{77}
\end{array} \right)= 
\left( \begin{array}{ccccc}
c^2\left(e^{-2\phi} + G_{11} \Psi_1^2 \right) & 0  & 0 & \cdots & 0\\
0 & G_{22} &
0 & \cdots & 0 \\
0 & 0 & {G}_{33} & 0 \cdots & \vdots \\
\vdots & & & \ddots & \\
0& & \cdots  & 0 & {G}_{77}
\end{array} \right)  \nonumber\\
\label{gdual}
\eeqa
as well as 
\beqa
e^{2 {\tilde \phi}} = c^2 \left(G^{11} + e ^{2 \phi} \Psi_1^2 \right)
 \;\;\;\;\;\;,\;\;\;\;\;\;
{\tilde \Psi}=
\left( \begin{array}{c}
{\tilde \Psi}_1 \\
{\tilde \Psi}_2  \\
\vdots \\
{\tilde \Psi}_7 \\
{\tilde \Psi}_8 \\
{\tilde \Psi}_9 \\
\vdots
\end{array} \right) = 
\left( \begin{array}{c}
-\left(\frac{a}{c} 
+ \frac{e^{2 \phi} \Psi_1}{c^2 ( G^{11} + e^{2 \phi} \Psi_1^2)}
\right)\\
0 \\
\vdots \\
0 \\
\Psi_8 \\
0\\
\vdots
\end{array} \right) \;\;\;.
\eeqa
Note that ${\tilde \phi}$ now depends on both $r$ and $\theta$.

Next, we would like to determine the Killing spinor 
associated with the soliton background (\ref{gdual}).
The Killing spinor equations (\ref{killing}) now take the form
\beqa
\delta \chi^I&=& 0 \;\;\;,   \no\\
\delta\lambda &=&-\half e^{-{\tilde \phi}}\del_\mu 
\{{\tilde \phi} + \ln \det {\tilde e}_m^a \}
\gamma^\mu \varepsilon
+\hhalf e^{-2{\tilde \phi}}
{\tilde e}^m_d {\tilde F}^{(2)}_{\mu\nu m}\gamma^{\mu\nu} \gamma^4
\otimes\Sigma^d
\varepsilon \;\;\;, 
\label{duallam2}\\
\delta \psi_\mu &=&\del_\mu\varepsilon + {1\over 4}\omega_{\mu\alpha\beta}
\gamma^{\alpha\beta}\varepsilon + \half  e_{\mu \alpha}
e_{\beta}^{\nu} \partial_{\nu} {\tilde
  \phi} \gamma^{\alpha \beta} 
\varepsilon \nonumber\\
 &-& {1\over 4}e^{-{\tilde \phi}}[{\tilde e}^m_a {\tilde F}^{(2)}_{\mu\nu(m)} 
- {\tilde e}_{ma} 
{\tilde F}^{(1)m}_{\mu\nu}]
\gamma^\nu \gamma^4 \otimes\Sigma^{a}\varepsilon \;\;\;, 
\label{dualgrav2}
\\
\delta\psi_d &=&-{1\over 2}e^{-{\tilde \phi}}{\tilde e}^m_d\del_\mu 
{\tilde e}_{md}
\gamma^\mu \gamma^4 \otimes\Sigma^{d}\varepsilon
-{1\over 8}e^{-2{\tilde \phi}}
[{\tilde e}^m_d {\tilde F}^{(2)}_{\mu\nu m} 
+ {\tilde e}_{md} {\tilde F}^{(1)m}_{\mu\nu}]
\gamma^{\mu\nu}\varepsilon \;\;\;.
\label{dualintgra2}
\eeqa
Note that in (\ref{dualintgra2}) there is no summation over $d$.

We will again take the Killing spinor $\varepsilon = \epsilon \otimes \chi$ 
to satisfy (\ref{cond1}) and (\ref{cond2}).  Hence, $\epsilon$
is given by (\ref{spinor1}).

Using that 
\beqa
e^{-2 {\tilde \phi}} {\tilde F}^{(1)}_{t r 1 } = - \frac{{\tilde G}^{11}}{R} 
  \partial_{\theta} {\tilde \Psi}_8 \;\;\;,\;\;\;
e^{-2 {\tilde \phi}} {\tilde F}^{(2)}_{t r 1 } = - \frac{{\tilde G}_{11}}{R}
  \partial_{\theta} {\tilde \Psi}_1 \;\;\;,\;\;\;
e^{-2 {\tilde \phi}} {\tilde F}^{(2)}_{t \theta 1 } = - {\tilde G}_{11} R 
  \partial_r {\tilde \Psi}_1 \;,
\eeqa
it can be checked that the Killing spinor
equation $\delta \lambda = 0$ is satisfied provided that 
\beqa
\Sigma^1 \chi = \chi \;\;\;,\;\;\; {\tilde p} = 
\eta_{\alpha_8} \;\;\;,
\label{condeps}
\eeqa
where $\eta_{\alpha_8} ={ \alpha_8 /|\alpha_8|}$ denotes the sign of the
charge $\alpha_8$.  
Similarly, it can be checked that the Killing spinor equation 
$\delta \psi_1=0$ (eq. (\ref{dualintgra2}))  is satisfied.

Next, consider solving the Killing spinor equations (\ref{dualgrav2}).
We will again make the ansatz that the Killing spinor is static, 
that is $\varepsilon = \varepsilon(r, \theta)$.  Then, the equation
$\delta \psi_t = 0$ is satisfied.
The condition $\delta \psi_r = 0$, on the other hand, results in
\beqa
\partial_r \log {\tilde \epsilon} = i \frac{p}{2a} \, \frac{1}{\sqrt{r}} \, 
\frac{e^{2 \phi} \Psi_1 \partial_{\theta} \Psi_1}{G^{11} + e^{2 \phi} 
\Psi^2_1} + \frac{\eta_{\alpha_8} \sqrt{G_{11}}}{2a} \frac{1}{r^2}
\, \frac{\Psi_1^2 \partial_{\theta} \Psi_1}{G^{11} + e^{2 \phi} \Psi^2_1} 
\;\;\;,
\label{epsthreer}
\eeqa
whereas the condition $\delta \psi_{\theta} = 0$ results in
\beqa
\partial_{\theta} \log {\tilde \epsilon} = - i \frac{p}{4} \, \frac{a \,
G^{11}}
{\sqrt{r}} \, 
\frac{1}{G^{11} + e^{2 \phi} 
\Psi^2_1} - \frac{\eta_{\alpha_8} a \sqrt{G^{11}}}{4} 
\, \frac{e^{2 \Phi}\Psi_1}{G^{11} + e^{2 \phi} \Psi^2_1} 
\;\;\;.
\label{epsthreet}
\eeqa
Clearly, the solution to both
(\ref{epsthreer}) and (\ref{epsthreet}) will be of the form
$\log {\tilde \epsilon} = X + i Y$ with real $X$ and $Y$, namely
\beqa
{\tilde \epsilon} = e^{{\tilde \phi}/{2}} e^{i Y} \;\;\;,\;\;\;
Y = - \frac{\eta_{\alpha_8} p}{2} \; \arctan \left(
\frac{\eta_{\alpha_8} a}{2} \, \frac{\theta}{\sqrt{r}} \right) 
,
\eeqa
up to a multiplicative constant.  Comparison with (\ref{elspinor}) shows that,
whereas the form of $X$ was to be expected on the grounds of the 
replacement $ \phi \rightarrow {\tilde \phi}$ under duality,
the duality transformation ${\cal M}  \rightarrow
{\tilde {\cal M}} = \Omega {\cal M} \Omega^T$ 
actually also produces a complicated phase $Y$.

Note that when $r\rightarrow\infty$, the Killing spinor approaches a
constant value given by 
\eq
 {\tilde    \epsilon} \rightarrow (c^2 \,G^{11})^\hhalf\,
e^{\pm i( {n\over 2}\pi)}.
\en

This solitonic solution preserves $1/2$ of $N=8$ supersymmetry.

\subsection{Soliton solutions preserving $N=2$ supersymmetry}

In this subsection, we will consider 
soliton solutions preserving 
$1/4$ of $N=8, D=3$ supersymmetry.

A particular class
of such solutions can be obtained by combining certain features 
of the electrically charged solutions, discussed in subsection \ref{2c}, 
and of the solitonic solution discussed in subsection \ref{n41}.
Namely, we will make the following ansatz for the background fields
$G^{-1}$ and $B$,
\renewcommand{\arraystretch}{0.6}
\beqa
\label{gbdyon}
\left(\!\!\! \begin{array}{cccccc}
{G}^{11} & {G}^{12}& 0 & 0 &  \cdots & 0\\
{G}^{21} & {G}^{22}& 0 & 0 & \cdots & 0 \\
0 & 0 & {G}^{33} & 0 & \cdots    &  \\
0 & 0 & 0 &  {G}^{44} & 0  &  \vdots \\ 
\vdots &  &   &  & \ddots &   \\
0& & \cdots &  & 0 & {G}^{77}
\end{array}\!\!\! \right)
\!\!=\!\!
\left(\!\!\!\! \begin{array}{cccccc}
f^2(r) & - f^2(r) \Upsilon_2  & 0 &0& \cdots & 0\\
- f^2(r) \Upsilon_2 & |\frac{\alpha_9}{\alpha_2}| 
+  f^2(r) \Upsilon_2^2 & 0& 
0 & \cdots & 0 \\
0 & 0 & {G}^{33} & 0 &  \cdots &  \\
0 & 0 & 0 &  {G}^{44} & 0  & \vdots  \\ 
\vdots & & & &  \ddots  \\
0& & \cdots  &  & 0& {G}^{77}
\end{array}\!\!\! \right)\!, &&\nonumber\\
\no\\
\renewcommand{\arraystretch}{1.0}
{ B} =  ({ B}_{mn}) = 
\left( \begin{array}{ccccc}
0 &{B}_{12} & 0 & \cdots &0\\
{ B}_{21} & 0 \\
0& & 0 & & \vdots\\
\vdots& & & \ddots \\
0& &\cdots & & 0
\end{array} \right)= 
\left( \begin{array}{ccccc}
0 &  \Upsilon_9 & 0 & \cdots &0\\
- \Upsilon_9 & 0 \\
0& & 0 & & \vdots\\
\vdots& & & \ddots \\
0& &\cdots & & 0
\end{array} \right) ,\qquad\qquad\qquad &&
\eeqa
where
\renewcommand{\arraystretch}{1.0}

\beqa
f(r) = D \; r^{-\frac{\gamma}{2}} \;\;\;,\;\;\;
\Upsilon_2 = - \frac{\theta}{2 \pi} \alpha_9 \;\;\;, \;\;\;
\Upsilon_9 = - \frac{\theta}{2 \pi} \alpha_2 \;\;\;,\;\;\;
G^{44} = |\frac{\alpha_{11}}{\alpha_4}| .
\eeqa
We will also take
\beqa
e^{2  \phi} &=& r^{-\beta} \;\;\;\;\;\;,\;\;\;\;\;\;
{ \Psi}=
\left( \begin{array}{c}
{\Psi}_1 \\
{\Psi}_2 \\
{ \Psi}_3 \\
{\Psi}_4 \\
\Psi_5 \\
\vdots \\
{ \Psi}_{10} \\
{ \Psi}_{11} \\
{ \Psi}_{12} \\
\vdots
\end{array} \right) = 
\left( \begin{array}{c}
0\\
0\\
0\\
- \frac{\theta}{2 \pi} \alpha_{11} \\
0\\
\vdots \\
0\\
- \frac{\theta}{2 \pi} \alpha_4\\
0\\
\vdots
\end{array} \right) \;\;\;\;\;\;,\;\;\;\;\;\; {a}_m^I = 0 \;\;\;\;.
\eeqa
For the 
space--time metric we will make the ansatz
\beqa
ds^2 = -  dt^2 +  dr^2 + R^2 (r) \, d\theta^2 \;\;\;,\;\;\;
R(r) = a r^{\rho} \;\;\;.
\eeqa
The constants $D$, $\beta$, $\gamma$ and $\rho$ 
will be fixed below.

The internal inverse vielbein
${ e}_a^{m}$ associated to (\ref{gbdyon}) 
is given by
\beqa
{ e}_a^{m}&=&  
\left( \begin{array}{cccccc}
  f(r) & - f(r) \Upsilon_2
& 0 & \cdots & & 0\\
0 & \sqrt{|\frac{\alpha_9}{\alpha_2}|}& 0 & &  \cdots & 0 \\
0 & 0 & \sqrt{G^{33}} & 0 & \cdots & \\
0 & 0 & 0& \sqrt{|\frac{\alpha_{11}}{\alpha_4}|} & 0 & \vdots \\
\vdots & & & & \ddots & \\
0& & \cdots  & & 0 & \sqrt{{G}^{77}}
\end{array} \right) \;\;\;.
\eeqa

Next, we would like to determine the Killing spinor 
associated with the soliton background (\ref{gbdyon}).
The Killing spinor equations (\ref{killing}) now take the form
\beqa
\delta \chi^I&=& 0 \;\;\;,\;\;\; \nonumber\\
\delta\lambda &=& -\half e^{-\phi}\del_\mu \{\phi+\ln\det e_m^a\} 
\gamma^\mu\,\varepsilon 
+\hhalf e^{-2\phi} F^{(2)}_{\mu\nu m}
\gamma^{\mu\nu}\gamma^4\otimes\Sigma^m\varepsilon
\label{lamdy}\no\\
&&+\hhalf e^{-\phi}\del_\mu B_{mn}
\gamma^\mu\otimes\Sigma^{mn}
\varepsilon \;\;\;,\\
\delta{\psi}_\mu &=&\del_\mu\varepsilon + {1\over 4}\omega_{\mu\alpha\beta}
\gamma^{\alpha\beta}\varepsilon 
 +\hhalf (e_{\mu\alpha}e_\beta^\nu \!-\!e_{\mu\beta}e_\alpha^\nu)
\del_\nu\phi\gamma^{\alpha\beta}\varepsilon + {1\over 8}(e^n_a\del_\mu 
e_{nb}\!-\!e^n_b\del_\mu e_{na})\Sigma^{ab}\varepsilon \no\\
&&
-{1\over 4}e^{-\phi}[e^m_a F^{(2)}_{\mu\nu(m)} 
- e_{ma} F^{(1)m}_{\mu\nu}]\gamma^\nu\gamma^4\otimes\Sigma^a\varepsilon
-{1\over 8}\del_\mu B_{mn} 
\Sigma^{mn}\varepsilon 
\label{gravdy}\;\;\;,\\
\delta\psi_d &=&-{1\over 4}e^{-\phi}(e_d^m\del_\mu e_{ma}\!+\!e_a^m\del_\mu 
e_{md})\gamma^\mu\gamma^4\otimes\Sigma^a\varepsilon 
+{1\over 4}e^{-\phi}e^m_de^n_a \del_\mu B_{mn}\gamma^\mu
\gamma^4\otimes\Sigma^a\varepsilon\no\\
&&-{1\over 8}e^{-2\phi}[e_{md}F_{\mu\nu}^{(1)m}
+e_d^mF^{(2)}_{\mu\nu m}]\gamma^{\mu\nu}\varepsilon \;\;\;.
\label{ingravdy}
\eeqa
As before, 
the Killing spinor $\varepsilon = \epsilon \otimes \chi$ 
will be taken to satisfy (\ref{cond1}) and (\ref{cond2}) and, hence,
also (\ref{spinor1}).
The Killing spinor equation $\delta \lambda =0$ can be solved by demanding
that
\beqa
\Sigma^{12}  \chi = q\, i\, \chi 
 \;\;\;,\;\;\; \Sigma^4 \chi = \chi \;\;\;,\;\;\;\; q = \pm \;\;\;.
\label{dyc2}
\eeqa
Note that the condition (\ref{dyc2}) reduces the
degrees of freedom of $\varepsilon$ to 
$4$ real degrees of freedom, and thus
the solitonic background under consideration preserves $1/4$ of $N=8$
supersymmetry.
Then, the Killing spinor equation $\delta \lambda =0$ is solved provided that
\beqa
\beta = \gamma \;\;\;,\;\;\; \rho = 1 - \frac{\gamma}{2} \;\;\;,\;\;\;
D = \gamma \frac{\pi a}{\sqrt{|\alpha_2 \alpha_{9}|}} \;\;\;,\;\;\;
\pi a = \frac{1}{\beta} \sqrt{|\alpha_4 \alpha_{11}|}
\label{consdy}
\eeqa
as well as $q = -p \eta_{\alpha_2}$ and ${\tilde p} = 
\eta_{\alpha_{11}}$, where $\eta_{\alpha_2}$
and $\eta_{\alpha_{11}}$ denote the signs of the charges 
$\alpha_2$ and $\alpha_{11}$, respectively ($\eta_{\alpha_2}=-
\eta_{\alpha_9}, \eta_{\alpha_4}= -\eta_{\alpha_{11}}$).

Next, consider solving the Killing spinor equations (\ref{gravdy}).
We will again make the ansatz that the Killing spinor is static.  Then, the
equation $\delta \psi_{t} = 0$ is automatically satisfied.  The condition
$\delta \psi_{r} = 0$, on the other hand, yields
\beqa
\partial_r \log {\tilde \epsilon} = \frac{1}{2} \partial_r \phi \;\;\;.
\label{rdy}
\eeqa
Finally, the condition
$\delta \psi_{\theta} = 0$ can be solved by setting 
\beqa
\beta = 2(1-\gamma) \;\;\; .
\label{bg}
\eeqa
Then $\delta \psi_{\theta} = \partial_{\theta} \varepsilon
=0$ and, hence,
\beqa
{\tilde \epsilon} = e^{\frac{\phi}{2}} \;\;\;,
\eeqa
up to a multiplicative constant.  Comparison of (\ref{consdy}) and 
(\ref{bg}), on the other hand, yields that $\beta = \gamma = \rho= 2/3$.
Thus, it follows that
\beqa
f^2(r) &=& D^2 \, r^{- \frac{2}{3}} \;\;\;,\;\;\;
R(r) = a r^{\frac{2}{3}} \;\;\;,\;\;\;
e^{2  \phi} = r ^{- \frac{2}{3}} \;\;\;, \nonumber\\
D &=& \frac{\sqrt{|\alpha_4 \alpha_{11}|}}{
\sqrt{|\alpha_2 \alpha_{9}|}}\;\;\;,\;\;\;
a = \frac{3}{2 \pi} \sqrt{|\alpha_4 \alpha_{11}|} \;\;\;.
\eeqa
Then, finally, it can be checked that the Killing spinor equations
$\delta \psi_1=0$, $\delta \psi_2=0$ and 
$\delta \psi_4=0$ (eqs. (\ref{ingravdy})) are also satisfied.

Other solitonic solutions preserving 
$1/4$ of $N=8$
supersymmetry can be obtained by applying the $O(8,24)$ duality
transformation
\beqa
\Omega =  \left( \begin{array}{ccccccccc}
I_2 & 0  & 0 & 0 & 0 & 0 & 0 & 0  & 0 \\
0 & a & 0   & 0 & 0 & 0 & 0 & b & 0  \\
0 & 0 & I_3 & 0 & 0 & 0 & 0 & 0 & 0  \\
0 & 0  & 0 & I_2 & 0 & 0 & 0 & 0 & 0  \\
0 & 0 & 0   & 0 & d & 0 & 0 & 0 &  -c  \\
0 & 0 & 0  & 0 & 0 & I_3 & 0 & 0 & 0  \\
0 & 0 & 0  & 0& 0  & 0 & I_{16} & 0 & 0  \\
0 & c & 0 & 0 & 0 & 0 & 0 & d & 0  \\
0 & 0 & 0 & 0 & -b & 0 & 0 & 0 &  a
\end{array} \right) \;\;\;\;,\;\;\;\; ad-bc = 1
\eeqa
to (\ref{gbdyon}).  
We will, for concreteness, set $a = d= 0, b = -c = 1$
in the following.
The resulting dual background fields $\tilde{G}^{-1}$ and $\tilde{B}$ 
are then given by
\renewcommand{\arraystretch}{0.7}
\beqa
\left( \begin{array}{cccccc}
{{\tilde G}}^{11} & {{\tilde G}}^{12}& 0 & 0 &  \cdots & 0\\
{{\tilde G}}^{21} & {{\tilde G}}^{22}& 0 & 0 & \cdots & 0 \\
0 & 0 & {{\tilde G}}^{33} &{{\tilde G}}^{34}  & \cdots    &  \\
0 & 0 & {\tilde G}^{43} &  {\tilde G}^{44} & 0  &  \vdots \\ 
\vdots &  &   &  & \ddots &   \\
0& & \cdots &  & 0 & {\tilde G}^{77}
\end{array} \right)= 
\left( \begin{array}{cccccc}
G^{11} & G^{12}  & 0 &0& \cdots & \\
G^{21} & G^{22} & 0& 
0 & \cdots & 0 \\
0 & 0 & e^{2 \phi} & - e^{2 \phi}\Psi_4 &  \cdots &  \\
0 & 0 & - e^{2 \phi}\Psi_4 &  {G}^{44} + e^{2 \phi} \Psi_4^2 & 0  & \vdots \\ 
\vdots & & & &  \ddots  \\
0& & \cdots  &  & 0& {G}^{77}
\end{array} \right),&&\no\\
 \nonumber\\
{\tilde  B} =
\left( \begin{array}{cccccc}
0 &{\tilde B}_{12} & 0 & & \cdots &0\\
{ \tilde B}_{21} & 0 &  \\
0& & 0 & {\tilde B}_{34} & 0 &  \\
0& & {\tilde B}_{43}  & 0 & &  \vdots\\
\vdots& & 0 & &  \ddots \\
0& &\cdots & &&  0
\end{array} \right)= 
\left( \begin{array}{cccccc}
0 &  \Upsilon_9 & 0 & &  \cdots &0\\
- \Upsilon_9 & 0 \\
0& & 0 & \Psi_{11} & 0 & \\
0& & -  \Psi_{11} & 0 & &  \vdots\\
\vdots& &0 & &\ddots \\
0& &\cdots && & 0
\end{array} \right) ,\qquad\qquad\no &&
\eeqa
as well as 
\renewcommand{\arraystretch}{1.0}
\beqa
e^{2 {\tilde \phi}} = G^{33} \;\;\;,\;\;\; {\tilde \Psi} = 0 \;\;\;,\;\;\;
{\tilde a}_m^I = 0 \;\;\;.
\eeqa
Note that the dual dilaton field $\tilde \phi$ is constant.

It can be checked that the associated Killing spinor equations
are satisfied by a constant 
Killing spinor $\varepsilon = \epsilon \otimes \chi $
provided that
\beqa
\Sigma^{12}  \chi = - i\,p \, \eta_{\alpha_2} \chi 
 \;\;\;,\;\;\; \Sigma^{34} \chi = - i\,p \, \eta_{\alpha_4}
\chi \;\;\;, 
\eeqa
where, again, $\eta_{\alpha_2}$ and 
$\eta_{\alpha_4}$ denote the sign of the charges 
${\alpha_2}$ and ${\alpha_4}$, respectively.

\subsection{Soliton solutions preserving $N=1$ supersymmetry}

It is now straightforward to construct solutions which preserve
$1/8$ of $D=3, N=8$ supersymmetry.

One class of solitonic solutions preserving $1/8$ of $D=3, N=8$ supersymmetry
is given as follows.  The background fields are given by
\renewcommand{\arraystretch}{0.7}
\beqa
{ e}_a^{m}&=&  
\left( \begin{array}{cccccccc}
  f_2(r) & - f_2(r) \Upsilon_2
& 0 & 0 & 0 & 0 & 0\\
0 & \sqrt{|\frac{\alpha_9}{\alpha_2}|}& 0 & 0 & 0 & 0  & 0 \\
0&0 &f_4(r) & - f_4(r) \Upsilon_4 & 0 & 0 & 0 \\
0 & 0 &0  &\sqrt{|\frac{\alpha_{11}}{\alpha_4}|}&0&0&0 \\
0 & 0 & 0&0&\sqrt{G^{55}} & 0 & 0 \\
0&0 &0 &0 &0&\sqrt{G^{66}} &0 \\
0&0 & 0  &0 & 0 &0 &  \sqrt{{G}^{77}}
\end{array} \right) \;\;\;, \\
\no\\
 B &=&  
\left( \begin{array}{cccccccc}
0 &{ B}_{12} & 0 & 0 &0 &0&  0\\
{ B}_{21} & 0 & 0&0&0&0&0 \\
0&0 & 0  &B_{34}&0&0&0  \\
0& 0 & B_{43} &0 & 0 & 0&0\\
0&0 & 0 & 0& 0& 0 & 0  \\
0&0 & 0 & 0& 0& 0 & 0   \\
0&0 & 0 & 0& 0& 0 & 0\\
\end{array} \right)= 
\left( \begin{array}{cccccccc}
0 &\Upsilon_9 & 0 & 0 &0 &0&  0\\
-\Upsilon_9 & 0 & 0&0&0&0&0 \\
0&0 & 0  &\Upsilon_{11}&0&0&0  \\
0& 0 &- \Upsilon_{11} &0 & 0 & 0&0\\
0&0 & 0 & 0& 0& 0 & 0  \\
0&0 & 0 & 0& 0& 0 & 0   \\
0&0 & 0 & 0& 0& 0 & 0\\
\end{array} \right) \no
\eeqa
as well as 
\renewcommand{\arraystretch}{1.0}
\beqa
e^{2  \phi} &=& r^{-\gamma} \;\;\;\;\;\;,\;\;\;\;\;\;
{ \Psi}=
\left( \begin{array}{c}
{\Psi}_1 \\
\vdots \\
\Psi_6 \\
\Psi_7 \\
\Psi_8\\
\vdots\\
{ \Psi}_{13} \\
{ \Psi}_{14} \\
{\Psi}_{15}\\
\vdots
\end{array} \right) = 
\left( \begin{array}{c}
0\\
\vdots \\
0\\
- \frac{\theta}{2 \pi} \alpha_{14} \\
0\\
\vdots \\
0\\
- \frac{\theta}{2 \pi} \alpha_7\\
0\\
\vdots
\end{array} \right) \;\;\;\;\;\;,\;\;\;\;\;\; {a}_m^I = 0 \;\;\;\;.
\eeqa
Here
\beqa
\Upsilon_2 &=& - \frac{\theta}{2 \pi} \alpha_9 \;\;\;,\;\;\;
\Upsilon_4 = - \frac{\theta}{2 \pi} \alpha_{11} \;\;\;,\;\;\;
\Upsilon_9 = - \frac{\theta}{2 \pi} \alpha_{2} \;\;\;,\;\;\;
\Upsilon_{11} = - \frac{\theta}{2 \pi} \alpha_4 \;\;\;, \no\\
G_{77} &=& | \frac{\alpha_7}{\alpha_{14}}|\;\;\;
\eeqa
as well as 
\beqa
f_2 &=& D_2 \; r^{-\frac{\gamma}{2}} \;\;\;,\;\;\;
D_2=\frac{\sqrt{|\alpha_7 \alpha_{14}|}}{\sqrt{|\alpha_2 \alpha_9|}}\;\;\;,
\no\\
f_4 &=& D_4 \; r^{- \frac{\gamma}{2}}\;\;\;,\;\;\;
D_4=\frac{\sqrt{|\alpha_7 \alpha_{14}|}}{\sqrt{|\alpha_4 \alpha_{11}|}}\;\;\;.
\eeqa
The space--time metric is given by
\beqa
ds^2 = - dt^2 + dr^2 + R^2 d\theta^2 \;\;\;,\;\;\; R = a r^{1 
- \frac{\gamma}{2}} \;\;\;,
\eeqa
where
\beqa
a = \frac{1}{\gamma \, \pi} \sqrt{|\alpha_7 \alpha_{14}|} \;\;\;,\;\;\;
\gamma = \frac{1}{2} \;\;\;.
\eeqa
The associated Killing spinor $\varepsilon = \epsilon \otimes \chi$
satisfies (\ref{cond1}) and (\ref{cond2}) with
\beqa
{\tilde \epsilon} = e^{\frac{\phi}{2}} \;\;\;,
\eeqa
as well as
\beqa
\Sigma^{12} \chi = i q \chi \;\;\;,\;\;\;
\Sigma^{34} \chi = i {\tilde q} \chi \;\;\;,\;\;\;
\Sigma^{7} \chi = \chi \;\;\;,\;\;\; q = \pm \;\;\;,\;\;\; {\tilde q} = 
\pm \;, 
\eeqa
where ${\tilde p} = \eta_{\alpha_{14}}, q = - p \eta_{\alpha_2}$
and ${\tilde q} = - p \eta_{\alpha_4}$ ($\eta_{\alpha_2}
= - \eta_{\alpha_9}, \eta_{\alpha_4}= -  \eta_{\alpha_{11}}, 
\eta_{\alpha_7} = - \eta_{\alpha_{14}}$).

Another
class of solitonic solutions preserving $1/8$ of $D=3, N=8$ supersymmetry
is given as follows.  The background fields are given by
\renewcommand{\arraystretch}{0.7}
\beqa
&&{ e}_a^{m}=  
\left( \begin{array}{cccccccc}
  f_2(r) & - f_2(r) \Upsilon_2
& 0 & 0 & 0 & 0 & 0\\
0 & \sqrt{|\frac{\alpha_9}{\alpha_2}|}& 0 & 0 & 0 & 0  & 0 \\
0&0 &f_4(r) & - f_4(r) \Upsilon_4 & 0 & 0 & 0 \\
0 & 0 &0  &\sqrt{|\frac{\alpha_{11}}{\alpha_4}|}&0&0&0 \\
0 & 0 & 0&0&f_6(r)   &  - f_6(r) \Upsilon_6  & 0 \\
0&0 &0 &0 &0&\sqrt{|\frac{\alpha_{13}}{\alpha_6}|} &0 \\
0&0 & 0  &0 & 0 &0 &  \sqrt{{G}^{77}}
\end{array} \right) \;\;,\no\\ 
\no\\
 B =\!\!\!\!\!\!\!\!\!   
&&\left( \begin{array}{cccccccc}
0 &{ B}_{12} & 0 & 0 &0 &0&  0\\
{ B}_{21} & 0 & 0&0&0&0&0 \\
0&0 & 0  &B_{34}&0&0&0  \\
0& 0 & B_{43} &0 & 0 & 0&0\\
0&0 & 0 & 0& 0& B_{56}& 0  \\
0&0 & 0 & 0& B_{65}& 0 & 0   \\
0&0 & 0 & 0& 0& 0 & 0\\
\end{array} \right)\!\!=\!\! 
\left( \begin{array}{cccccccc}
0 &\Upsilon_9 & 0 & 0 &0 &0&  0\\
-\Upsilon_9 & 0 & 0&0&0&0&0 \\
0&0 & 0  &\Upsilon_{11}&0&0&0  \\
0& 0 &- \Upsilon_{11} &0 & 0 & 0&0\\
0&0 & 0 & 0& 0&\Upsilon_{13} & 0  \\
0&0 & 0 & 0& - \Upsilon_{13}& 0 & 0   \\
0&0 & 0 & 0& 0& 0 & 0\\
\end{array} \right) \no \\
\eeqa
\renewcommand{\arraystretch}{1.0}
as well as 
\beqa
e^{2  \phi} &=& r^{-\gamma} \;\;\;\;\;\;,\;\;\;\;\;\;
{ \Psi}=
\left( \begin{array}{c}
{\Psi}_1 \\
\vdots \\
\Psi_6 \\
\Psi_7 \\
\Psi_8\\
\vdots\\
{ \Psi}_{13} \\
{ \Psi}_{14} \\
{\Psi}_{15}\\
\vdots
\end{array} \right) = 
\left( \begin{array}{c}
0\\
\vdots \\
0\\
- \frac{\theta}{2 \pi} \alpha_{14} \\
0\\
\vdots \\
0\\
- \frac{\theta}{2 \pi} \alpha_7\\
0\\
\vdots
\end{array} \right) \;\;\;\;\;\;,\;\;\;\;\;\; {a}_m^I = 0 \;\;\;\;.
\eeqa
Here
\beqa
\Upsilon_2 &=& - \frac{\theta}{2 \pi} \alpha_9 \;\;\;,\;\;\;
\Upsilon_4 = - \frac{\theta}{2 \pi} \alpha_{11} \;\;\;,\;\;\;
\Upsilon_6 = - \frac{\theta}{2 \pi} \alpha_{13} \;\;\;,\;\;\; \no\\
\Upsilon_9 &=& - \frac{\theta}{2 \pi} \alpha_{2} \;\;\;,\;\;\;
\Upsilon_{11} = - \frac{\theta}{2 \pi} \alpha_4 \;\;\;,\;\;\;
\Upsilon_{13} = - \frac{\theta}{2 \pi} \alpha_6 \;\;\;,\no\\
G_{77} &=& | \frac{\alpha_7}{\alpha_{14}}|\;\;\;
\eeqa
as well as 
\beqa
f_2 &=& D_2 \; r^{-\frac{\gamma}{2}} \;\;\;,\;\;\;
D_2=\frac{\sqrt{|\alpha_7 \alpha_{14}|}}{\sqrt{|\alpha_2 \alpha_9|}}\;\;\;,
\no\\
f_4 &=& D_4 \; r^{-\frac{\gamma}{2}}\;\;\;,\;\;\;
D_4=\frac{\sqrt{|\alpha_7 \alpha_{14}|}}{\sqrt{|\alpha_4
    \alpha_{11}|}}\;\;\;,
\no\\
f_6 &=& D_6 \; r^{-\frac{\gamma}{2}}\;\;\;,\;\;\;
D_6=\frac{\sqrt{|\alpha_7 \alpha_{14}|}}{\sqrt{|\alpha_6 \alpha_{13}|}}\;\;\;.
\eeqa
The space--time metric is given by
\beqa
ds^2 = - dt^2 + dr^2 + R^2 d\theta^2 \;\;\;,\;\;\; R = a r^{1 
- \frac{\gamma}{2}} \;\;\;,
\eeqa
where
\beqa
a = \frac{1}{\gamma \, \pi} \sqrt{|\alpha_7 \alpha_{14}|} \;\;\;,\;\;\;
\gamma = \frac{2}{5} \;\;\;.
\eeqa
The associated Killing spinor $\varepsilon = \epsilon \otimes \chi$
satisfies (\ref{cond1}) and (\ref{cond2}) with
\beqa
{\tilde \epsilon} = e^{\frac{\phi}{2}} \;\;\;,
\eeqa
as well as
\beqa
\Sigma^{12} \chi = i q \chi \;,\;
\Sigma^{34} \chi = i {\tilde q} \chi \;,\;
\Sigma^{56} \chi = i {\hat q} \chi \;,\;
\Sigma^{7} \chi = \chi \;,\; q = \pm \;,\; {\tilde q} = 
\pm \;,\; {\hat q} = 
\pm ,
\eeqa
where ${\tilde p} = \eta_{\alpha_{14}}, q = - p \eta_{\alpha_2},
{\tilde q} = - p \eta_{\alpha_4}$ 
and ${\hat q} = - p \eta_{\alpha_6}$
($\eta_{\alpha_2}
= - \eta_{\alpha_9}, \eta_{\alpha_4}= -  \eta_{\alpha_{11}}, 
\eta_{\alpha_6}= -  \eta_{\alpha_{13}}, 
\eta_{\alpha_7} = - \eta_{\alpha_{14}}$).

\section{Supersymmetric solutions with $\vec{\alpha}^2=0$}

\setcounter{equation}{0}

In this section, we will consider a particular class of solutions to the
Killing spinor equations, namely solutions for which
$\vec{\alpha}^2=\alpha^TL\alpha = 0$. We will construct solutions which
preserve $1/2^m$ of $N=8, D=3$ supersymmetry, where $m=1,2,3$.
The solutions are obtained with $H_{\mu\nu\rho}=0$ and $a_m^I=0$.

We will find that the space--time metric (\ref{gg}) is given in terms of
\beqa
V=1,\qquad\qquad R = a \,r^{1 - {\gamma}} 
\;\;\;,
\eeqa
and that the dilaton is given by
\beqa
e^{2 \phi} = r^{-\gamma} \;\;\;,
\eeqa
where
\beqa
\gamma = \frac{2}{n+2} \;\;\;\;,\;\;\;\; 
a = \frac{ |\alpha_i|}{2 \pi \, \gamma} 
\;\;\;.
\eeqa
By the coordinate transformation $r=({\gamma})^{1\over\gamma}\;(a\ln
  \tilde{r})^{1\over\gamma}\,,\,
1 \leq {\tilde r} \leq {\infty}$,
the associated space--time metric can be put
  into the form
\eq
ds^2=-dt^2+ a^{2\over\gamma}\;({\gamma})^{{2(1-\gamma)}\over\gamma}\;
{{(\ln\tilde{r})}\over{\tilde{r}^2}}^{{2(1-\gamma)}\over\gamma}
(d\tilde{r}^2+\tilde{r}^2d\theta^2)\label{met1}.
\en
The curvature scalar, ${\cal R}=g^{\mu\nu}{\cal R}_{\mu\nu}$, 
is computed to be
\eq
{\cal R}\,=\,2\gamma(1-{\gamma})\;{1\over
  {r^2}}\,=\,{{4n}\over{(n+2)^2}}\,{1\over {r^2}}.
\en

\subsection{Electrically charged  solutions\label{1c}}

We will be solving the same Killing spinor equations subject to the same
assumptions  as in subsection \ref{2c}, where in addition we take
$\alpha_9=0$. 

Looking back at (\ref{lamb}), with $\alpha_9=0,\,$ i.e. $F_{\mu\nu
  2}^{(2)}=0$,   
we have $\, \del_r\phi=-\del_r\ln \sqrt{G_{22}}.\, $  {From}
(\ref{psitheta1}) we still have $\del_r\phi=-\del_r \ln R$.

These  relations can be satisfied with the following ansatz
\eq
\sqrt{G_{22}}=de^{-\phi},\qquad\qquad R=a\,e^{-\phi},
\en
where $d$ is an integration constant that is set to one in the following.

We can now solve  equations (\ref{psit}), (\ref{psid}) and find, (with 
again $V=1$),
\eqn
\sqrt{G_{22}}&=&r^{1\over 3},\no\\
R&=&{a\over c}\,r^{1\over 3},\qquad\qquad\qquad 
 {a\over {c^3}}=-{{3\,\tilde{p}\,\alpha_2}\over {4\pi}}\;\;,\no\\
e^{2\phi}&=& c^2r^{-{2\over 3}},
\enq
where $\tilde{p}\alpha_2=-|\alpha_2|$ and where the integration constant $c$  
will be set to one.

The space--time metric is now of the  form
\eq
ds^2=-dt^2+dr^2+ a^2\,r^{2\over
  3}\,d\theta^2.\label{metric1}
\en
or, equivalently,
\eq
ds^2= - dt^2 + {{2a^3}\over 3}\left({{\ln \tilde{r}}
\over {\tilde {r}^2}}\right)(d\tilde{r}^2 + \tilde{r}^2d\theta^2)\,,
\en
where $r={({{2a}\over 3})}^{3/2}(\ln {\tilde r})^{3/2}.$

The dependence of the spinor in terms of $\phi$ is the same as before
\eqn
&&\del_r \,\,\tilde{\epsilon}
- \half \del_r \phi \,\,\tilde{\epsilon}=0\;\;,\no\\
&&\del_\theta \,\,\tilde{\epsilon}=0\;\;,
\enq
or
\eq
\tilde{\epsilon}\,=\,e^{{\phi}\over 2}\, \,=\,
r^{-{1\over 6}}\;\;.
\en
These electric  solutions preserve again 1/2 of the $N=8$ supersymmetry.

\subsection{ Soliton solutions preserving $N=4$ supersymmetry}

Now we  discuss the soliton solution which is obtained by dualizing 
the charged solution discussed in subsection \ref{1c}, 
with one electric charge
only ($\alpha_9=0$). The  bosonic background fields of the charged solution
are given by $\phi, (G_{mn}) = {\rm diagonal} 
\left( {G}_{11}, G_{22}, \dots, {G}_{77}\right), G_{22} = r^{2\over 3},
B_{mn} = 0, a_m^I = 0$
and $\Psi^T = ( 0 , 0, 0 , \dots, 0, \Psi_9, 0, \dots, 0)\,=\,
( 0 ,\, 0\, , 0 , \dots, 0,\,-{\theta\over{2\pi}}\,\alpha_2 \,, 0, \dots, 0).$
We will utilize the $O(8,24)$ transformation $\Omega$ given in 
(\ref{omega}) to generate the dual background 
${\cal M}\rightarrow {\Omega}{\cal  M}{\Omega}^T$. We will for simplicity
set the transformation parameter $d$ to $d=0$ in the following so that
$bc=-1.$ 
The dual background fields are then given by
\beqa
{\tilde G}^{-1} &=&  
\left( \begin{array}{ccccc}
{\tilde G}^{11} & 0& 0 & \cdots & 0\\
0 & {\tilde G}^{22}& 0 & \cdots & 0 \\
0 & 0 & {\tilde G}^{33} & 0 \cdots & \vdots \\
\vdots & & & \ddots & \\
0& & \cdots  & 0 & {\tilde G}^{77}
\end{array} \right)= 
\left( \begin{array}{ccccc}
b^2 e^{2 \phi} & 0  & 0 & \cdots & 0\\
0 & G^{22}  &
0 & \cdots & 0 \\
0 & 0 & {G}^{33} & 0 \cdots & \vdots \\
\vdots & & & \ddots & \\
0& & \cdots  & 0 & {G}^{77}
\end{array} \right) \;, \nonumber\\
{\tilde B} &=&  ({\tilde B}_{mn}) = 
\left( \begin{array}{ccccc}
0 &{\tilde B}_{12} & 0 & \cdots &0\\
{\tilde B}_{21} & 0 \\
0& & 0 & & \vdots\\
\vdots& & & \ddots \\
0& &\cdots & & 0
\end{array} \right)= 
\left( \begin{array}{ccccc}
0 & - c \Psi_9 & 0 & \cdots &0\\
c \Psi_9 & 0 \\
0& & 0 & & \vdots\\
\vdots& & & \ddots \\
0& &\cdots & & 0
\end{array} \right) 
\label{gbdual1}
\eeqa
as well as 
\beqa
e^{2 {\tilde \phi}} &=& c^2 G^{11} \;\;\;\;\;\;,\;\;\;\;\;\;
{\tilde \Psi}=
\left( \begin{array}{c}
{\tilde \Psi}_1 \\
{\tilde \Psi}_2 \\
\vdots \\
{\tilde \Psi}_9 \\
\vdots
\end{array} \right) = 
\left( \begin{array}{c}
-a/c\\
0 \\
\vdots \\
0 \\
\vdots
\end{array} \right) \;\;\;\;\;\;,\;\;\;\;\;\; {\tilde a}_m^I = 0 \;\;\;\;.
\eeqa
The associated gauge field strengths $F_{\mu \nu}^{(a)}$
are again all zero for this solitonic solution.  The internal inverse vielbein
${\tilde e}_a^{m}$ associated to (\ref{gbdual1}) 
is given by
\beqa
{\tilde e}_a^{m}&=&  
\left( \begin{array}{ccccc}
 b e^{\phi} & 0
& 0 & \cdots & 0\\
0 & \sqrt{{G}^{22}}& 0 & \cdots & 0 \\
0 & 0 & \sqrt{{G}^{33}} & 0 \cdots & \vdots \\
\vdots & & & \ddots & \\
0& & \cdots  & 0 & \sqrt{{G}^{77}}
\end{array} \right) \;\;\;.
\eeqa

Note that the space--time metric is duality invariant and hence given as
in (\ref{metric1}).

The Killing spinor equations in the new background (\ref{gbdual1}) are of
the same form as in equations (\ref{duallam}),
(\ref{dualgrav}) and (\ref{dualintgrav}).  
It is easy to check that the Killing spinor will be of the same form as
before with the same conditions (\ref{cond1}), (\ref{cond2}) 
 and (\ref{dualchi}) to be satisfied.

For the solitonic background under consideration, the Killing spinor
equation (\ref{duallam}) then yields
\beqa
- p \,\partial_r \log \det {\tilde e}_m^a 
+b\, q\, 
\frac{\sqrt{G^{22}}}{R} e^{\phi}\partial_{\theta} {\tilde B}_{12} 
= 0 \;\;\;,
\eeqa
which is  satisfied, provided one takes $q = p\tilde{p}$.

Next, consider solving the Killing spinor equations (\ref{dualgrav}).
The Killing spinor being  static, the equation
$\delta \psi_t = 0$ is again automatically satisfied.  
The condition $\delta \psi_r =0$ yields again 
\beqa
\partial_r \varepsilon = 0 \;\;\;.
\eeqa
Finally, the condition $\delta \psi_{\theta} =0$
results in 
\beqa
\partial_{\theta} \varepsilon - \frac{1}{2} \partial_r R \gamma^{12} 
\varepsilon - \hhalf \sqrt{G^{22}} e^{\phi}
\partial_{\theta} \Psi_9 \Sigma^{12} \varepsilon =0 \;\;\;,
\label{prgrav}
\eeqa
from which it follows again, if $p\tilde{p}=q$, that
\beqa
\partial_{\theta} \varepsilon =0 \;\;\;.
\eeqa
Hence the Killing spinor $\varepsilon$ is constant.

Finally, it can be checked that the Killing spinor equations 
(\ref{dualintgrav}) for $\delta \psi_1 $ and $\delta \psi_2$
are automatically satisfied.

The solitonic background preserves 1/2 of $N=8$ supersymmetry.

\subsection{Soliton solutions preserving $N=2$ supersymmetry}

We take the following ansatz for the background fields $G^{-1}$ and $B$
\renewcommand{\arraystretch}{0.5}
\beqa
\left( \begin{array}{cccccc}
{G}^{11} & 0& 0 & 0 &  \cdots & 0\\
0 & {G}^{22}& 0 & 0 & \cdots & 0 \\
0 & 0 & {G}^{33} & 0 & \cdots    &  \\
0 & 0 & 0 &  {G}^{44} & 0  &  \vdots \\ 
\vdots &  &   &  & \ddots &   \\
0& & \cdots &  & 0 & {G}^{77}
\end{array} \right)=\left( \begin{array}{cccccc}
f_1^2(r) & 0  & 0 &0& \cdots & 0\\
0 &f_2^2(r)  & 0& 
0 & \cdots & 0 \\
0 & 0 & {G}^{33} & 0 &  \cdots &  \\
0 & 0 & 0 &f_4^2(r)  & 0  & \vdots  \\ 
\vdots & & & &  \ddots  \\
0& & \cdots  &  & 0& {G}^{77}
\end{array} \right),&& \nonumber\\
\no\\
{ B} =  ({ B}_{mn}) = 
\left( \begin{array}{ccccc}
0 &{B}_{12} & 0 & \cdots &0\\
{ B}_{21} & 0 \\
0& & 0 & & \vdots\\
\vdots& & & \ddots \\
0& &\cdots & & 0
\end{array} \right)= 
\left( \begin{array}{ccccc}
0 &  \Upsilon_9 & 0 & \cdots &0\\
- \Upsilon_9 & 0 \\
0& & 0 & & \vdots\\
\vdots& & & \ddots \\
0& &\cdots & & 0
\end{array} \right),\qquad\quad &&
\eeqa
where
\renewcommand{\arraystretch}{1.0}
\beqa
f_i^2(r) = \frac{D_i^2}{r^{\gamma}} \;\;\;,\;\;\;D_4=1,\;\;
\Upsilon_9 = - \frac{\theta}{2 \pi} \alpha_2 \;\;.
\eeqa
We will also take
\beqa
e^{2  \phi} &=& r^{-\gamma} \;\;\;\;\;\;,\;\;\;\;\;\;
{ \Psi}=
\left( \begin{array}{c}
{\Psi}_1 \\
{\Psi}_2 \\
{ \Psi}_3 \\
{\Psi}_4 \\
\Psi_5 \\
\vdots \\
{ \Psi}_{10} \\
{ \Psi}_{11} \\
{ \Psi}_{12} \\
\vdots
\end{array} \right) = 
\left( \begin{array}{c}
0\\
0\\
0\\
0 \\
0\\
\vdots \\
0\\
- \frac{\theta}{2 \pi} \alpha_4\\
0\\
\vdots
\end{array} \right) \;\;\;\;\;\;,\;\;\;\;\;\; {a}_m^I = 0 \;\;\;\;.
\eeqa
For the 
space--time metric we will make the ansatz
\beqa
ds^2 = -  dt^2 +  dr^2 + R^2 (r) \, d\theta^2 \;\;,\;\;\;\qquad
R(r) = a\; r^{1-\gamma} \;\;\;.
\eeqa
The constants $D_i$ and $\gamma$ will be fixed by the Killing spinor 
equations.

The Killing spinor equations (\ref{killing}) now take the form
\beqa
\delta \chi^I&=& 0 \;\;\;,\;\;\; \nonumber\\
\delta\lambda &=& -\half e^{-\phi}\del_\mu \{\phi+\ln\det e_m^a\} 
\gamma^\mu\,\varepsilon +\hhalf e^{-\phi}\del_\mu B_{mn}
\gamma^\mu\otimes\Sigma^{mn}
\varepsilon\label{lamb2} \;\;\;,\\
\delta{\psi}_\mu &=&\del_\mu\varepsilon + {1\over 4}\omega_{\mu\alpha\beta}
\gamma^{\alpha\beta}\varepsilon 
+\hhalf (e_{\mu\alpha}e_\beta^\nu \!-\!e_{\mu\beta}e_\alpha^\nu)
\del_\nu\phi\gamma^{\alpha\beta}\varepsilon\no\\
&&+\hhalf e^{-\phi}\;e_{ma} F^{(1)m}_{\mu\nu}\;
\gamma^\nu\gamma^4\otimes\Sigma^a\varepsilon 
-{1\over 8}\del_\mu B_{mn} \Sigma^{mn}\varepsilon 
\label{gravdy1}\;\;\;,\\
\delta\psi_d &=&-{1\over 4}e^{-\phi}(e_d^m\del_\mu e_{ma}\!+\!e_a^m\del_\mu 
e_{md})\gamma^\mu\gamma^4\otimes\Sigma^a\varepsilon 
+{1\over 4}e^{-\phi}e^m_de^n_a \del_\mu B_{mn}\gamma^\mu
\gamma^4\otimes\Sigma^a\varepsilon\no\\
&&-{1\over 8}e^{-2\phi}\;e_{md}F_{\mu\nu}^{(1)m}\;\gamma^{\mu\nu}
\varepsilon \;\;\;.
\label{ingravdy1}
\eeqa
The Killing spinor $\varepsilon=\epsilon\otimes\chi$ will be taken to
satisfy (\ref{cond1}) and (\ref{cond2}) as well as
$\Sigma^{12}\,\chi=iq\,\chi$ and $\Sigma^{4}\,\chi=\,\chi$.

The condition $\delta \lambda = 0$ (\ref{lamb2}) then implies
\eq
D_1D_2=-{{pq\gamma 2\pi a }\over {\alpha_2}}\;\;,
\en
while $\delta\psi_{\theta}=0$ in (\ref{gravdy1}) gives the condition
\eq
pa(1-{3\over 2}\gamma)+{{q\alpha_2}\over{4\pi}}D_1D_2=0\;\;.
\en
These last two equations yield $\gamma=\half$, 
whereas $\delta\psi_4=0$ (\ref{ingravdy1}) gives the relation
\eq
a=-{{\alpha_4}\over {2\pi\gamma\tilde{p}}}\;\;.
\en
Since this last quantity is positive, this implies
$\tilde{p}=-\eta_{\alpha_4}$, where $\eta_{\alpha_4}$ denotes the sign of
$\alpha_4$. 

The above shows that we can then take
\eq
D_1=D_2=\sqrt{-{{pq\gamma 2\pi a}\over \alpha_2}}=\sqrt{{\gamma 2\pi
    a}\over {|\alpha_2|}}\;\;,
\en
with $-pq=\eta_{\alpha_2}$.

This solution preserves 1/4 of $N=8$ supersymmetry.

\subsection{Soliton solutions preserving $N=1$ supersymmetry}

It is now straightforward to construct solutions which preserve
$1/8$ of $D=3, N=8$ supersymmetry.

One class of solitonic solutions preserving $1/8$ of $D=3, N=8$ supersymmetry
is given as follows.  The background fields $G^{-1}$ and $B$ are given by
\renewcommand{\arraystretch}{0.7}
\beqa
\left(\!\!\! \begin{array}{ccccccc}
{G}^{11} & 0& 0 & 0 & 0 & 0  & 0\\
0 & {G}^{22}& 0 & 0 & 0 & 0 & 0 \\
0 & 0 & {G}^{33} & 0 & 0 & 0  & 0 \\
0 & 0 & 0 &  {G}^{44} & 0  &  0 & 0\\ 
0 & 0 & 0  & 0 & G^{55} &0 & 0  \\
0 & 0 & 0 & 0 & 0 & G^{66} & 0\\
0 & 0 & 0 & 0 & 0 & 0 & {G}^{77}
\end{array}\!\!\! \right)\!\!=\!\!\left( \!\!\!\begin{array}{ccccccc}
f_1^2(r) & 0& 0 & 0 & 0 & 0  & 0\\
0 &f_2^2(r) & 0 & 0 & 0 & 0 & 0 \\
0 & 0 & {G}^{33} & 0 & 0 & 0  & 0 \\
0 & 0 & 0 & f_4^2(r)  & 0  &  0 & 0\\ 
0 & 0 & 0  & 0 &f_5^2(r)  &0 & 0  \\
0 & 0 & 0 & 0 & 0 &f_6^2(r)  & 0\\
0 & 0 & 0 & 0 & 0 & 0 & {G}^{77}
\end{array}\!\!\! \right),&& \no\\
\no\\
\left( \begin{array}{ccccccc}
0 & {B}_{12}& 0 & 0 & 0 & 0  & 0\\
{B}_{21} & 0 & 0 & 0 & 0 & 0 & 0 \\
0 & 0 & 0 & 0 & 0 & 0  & 0 \\
0 & 0 & 0 & 0  & 0  &  0 & 0\\ 
0 & 0 & 0  & 0 &0  &{B}_{56} & 0  \\
0 & 0 & 0 & 0 & {B}_{65} &0  & 0\\
0 & 0 & 0 & 0 & 0 & 0 & 0
\end{array} \right)
\!=\! \left( \begin{array}{ccccccc}
0 & \Upsilon_9& 0 & 0 & 0 & 0  & 0\\
-\Upsilon_9 & 0 & 0 & 0 & 0 & 0 & 0 \\
0 & 0 & 0 & 0 & 0 & 0  & 0 \\
0 & 0 & 0 & 0  & 0  &  0 & 0\\ 
0 & 0 & 0  & 0 &0  &\Upsilon_{13} & 0  \\
0 & 0 & 0 & 0 & -\Upsilon_{13} &0  & 0\\
0 & 0 & 0 & 0 & 0 & 0 & 0
\end{array} \right) ,\qquad\qquad &&\no
\eeqa
where
\renewcommand{\arraystretch}{1.0}
\beqa
f_i^2(r) = \frac{D_i^2}{r^{\gamma}} \;\;\;,\;\;\;D_4=1,\;\;
\Upsilon_9 = - \frac{\theta}{2 \pi} \alpha_2 \;\;\;, \;\;\;
\Upsilon_{13} = - \frac{\theta}{2 \pi} \alpha_6\;.
\eeqa
Again, we have
\beqa
e^{2  \phi} &=& r^{-\gamma} \;\;\;\;\;\;,\;\;\;\;\;\;
{ \Psi}=
\left( \begin{array}{c}
{\Psi}_1 \\
{\Psi}_2 \\
{ \Psi}_3 \\
{\Psi}_4 \\
\Psi_5 \\
\vdots \\
{ \Psi}_{10} \\
{ \Psi}_{11} \\
{ \Psi}_{12} \\
\vdots
\end{array} \right) = 
\left( \begin{array}{c}
0\\
0\\
0\\
0 \\
0\\
\vdots \\
0\\
- \frac{\theta}{2 \pi} \alpha_4\\
0\\
\vdots
\end{array} \right) \;\;\;\;\;\;,\;\;\;\;\;\; {a}_m^I = 0 \;\;\;\;.
\eeqa
As before, we will make the following ansatz for the   space--time metric 
\beqa
ds^2 = -  dt^2 +  dr^2 + R^2 (r) \, d\theta^2 \;\;\;,\;\;\;
R(r) = a\; r^{1-\gamma} \;\;\;.
\eeqa
Now, the Killing spinor equations give the following constraints
\eqn
&&-2\gamma p-{{\alpha_2q}\over {2\pi a}}D_1D_2 -
{{\alpha_6\tilde{q}}\over {2\pi a}}D_5D_6=0,\no\\
&&pa(1-{3\over 2}\gamma)+{{q\alpha_2}\over {4\pi}}D_1D_2 + 
{{\tilde{q}\alpha_6}\over {4\pi}}D_5D_6 =0,\label{d5d6}
\enq
as well as
\eq
D_1D_2=-pq\,{{2\pi a\gamma}\over{\alpha_2}},\qquad 
D_5D_6=-p\tilde{q}\,{{2\pi a\gamma}\over{\alpha_6}}\qquad{\rm and}\qquad 
 a=-{{\alpha_4\tilde{p}}\over{2\pi\gamma}}\;\;.
\en
Here we have used the conditions 
\eqn
\Sigma^4\chi&=&\chi\;\;,\no\\
\Sigma^{12}\chi&=&iq\chi,\qquad\qquad q=-p\eta_{\alpha_2}\;\;,\no\\
\Sigma^{56}\chi&=&i\tilde{q}\chi,\qquad\qquad
\tilde{q}=-p\,\eta_{\alpha_6}\;\;, 
\enq
which shows that the background preserves $1/8$ of the $N=8$ supersymmetry.

Note that equation (\ref{d5d6}) yields that $\gamma$ here is $2\over 5$.

Looking back at $B_{mn}$ we notice now that we can add one more block with
$B_{37}=-B_{73}=\Upsilon_{14}=-{\theta\over{2\pi}}\,\alpha_7$. 
The Killing equations will imply a further
condition on $\chi$, 
\eq
\Sigma^{37}\chi=i\hat{q}\,\chi,\qquad\qquad\hat{q}=-p\,\eta_{\alpha_7} \,\,.
\label{acon}
\en
Furthermore $\gamma=1/3$ and $D_3D_7=-p\hat{q}\,{{2\pi a\gamma}\over
  {\alpha_7}}$.  We note that the condition (\ref{acon}) does not
break any additional supersymmetry, and hence this solution also preserves
$1/8$ of the $N=8$ supersymmetry.

\section{The compactified cosmic string solution \label{cosmic1}}

\setcounter{equation}{0}

In \cite{sen1}, Sen constructed a particular three-dimensional 
 solution  by considering the fundamental string solution of the 
four-dimensional theory \cite{dab} and by winding the 
direction along which the string extends once in the third direction.  
As  shown in \cite{sen1}, this solution is related to the cosmic string
 solution of \cite{greene} by a O(8,24;$\IZ$) transformation.
Other three-dimensional solutions with internal winding 
can be obtained from the 
four-dimensional string solutions given in \cite{dfkr}.

The fundamental string solution in four space--time dimensions
is known to have partial space--time supersymmetry \cite{sen3}.
Here,
we will presently construct the Killing spinor associated with 
the particular three-dimensional  solution mentioned above.

The field configuration representing a fundamental string solution winding
once in the third direction is described below, following \cite{sen1}.

The three-dimensional space--time metric is now of the form
\eqn
ds^2&=&-dt^2+\lambda_2\,\bigl( dr^2+r^2\, d\theta^2\bigr)\no\\
&=&-dt^2+\lambda_2\,dzd{\bar{z}}\;\;,
\enq
where $z=re^{i\theta}$ is the complex coordinate labelling the two-dimensional
space. The scalar fields are
$e^{-2\phi}=\lambda_2,\; \Psi\equiv\Psi^1=-\lambda_1$
and $G_{11}=e^{2\phi}.$
{From} (\ref{psi}), the only non-vanishing field strength is
\eq
F_{\mu\nu1}^{(2)}= e^{4\phi}\epsilon^{\alpha\beta\gamma}\,
  e_{\mu\alpha}\,e_{\nu\beta}\,e_\gamma^\rho\del_\rho\Psi\;\;. 
\en
It will turn out to be convenient to combine $\lambda_1$ and $\lambda_2$
into a complex scalar field $S=\lambda_1 + i\lambda_2$. 
Here, as opposed to the previous cases, we take
$\Psi$ and $\phi$  to depend on both $r$ and  $\theta$.

With these assignments the Killing spinor equations take the form (where now 
$\Sigma^1\chi=\chi)$, 
\eqn
\delta\lambda &=&-(\del_r\phi\,\gamma^1 + {1\over
  r}\del_\theta\phi\,\gamma^2)\,\varepsilon +\half
  e^{2\phi}(\del_r\Psi\,\gamma^1 + {1\over
  r}\del_\theta\Psi\gamma^2 )\,J\,\gamma^4\,\varepsilon\;\;,\no\\
\delta{\psi}_t&=&-\half e^\phi (-\del_r\phi\,\gamma^2+{1\over
  r}\del_\theta\phi\,\gamma^1)\,J\,\varepsilon +\hhalf
e^{3\phi}(-\del_r\Psi\,\gamma^2+{1\over
  r}\del_\theta\Psi\,\gamma^1)\gamma^4\varepsilon\;\;,\no\\
\delta{\psi}_r&=&\del_r\,\varepsilon-\hhalf {{e^{2\phi}}\over r}
 \del_\theta\Psi\,\gamma^0\gamma^4\varepsilon\;\;,\no\\
\delta{\psi}_\theta &=&\del_\theta\,\varepsilon-\half\gamma^{12}
\,\varepsilon+{r\over 4}\,e^{2\phi}\del_r\Psi\,\gamma^0
\gamma^4\varepsilon\;\;,\no\\ 
\delta\psi_1&=&-\half(\del_r\phi\,\gamma^1+{1\over
  r}\del_\theta\phi\,\gamma^2)\,\gamma^4\,\varepsilon -\hhalf
  e^{2\phi}(\del_r\Psi\,\gamma^1 +{1\over
  r}\del_\theta\Psi\,\gamma^2)\,J\,\varepsilon\;\;.
\enq
Compatibility  of the form 
of $\epsilon$ for $\delta\lambda, 
\delta{\psi}_t$ and $\delta\psi_1$ can be achieved by imposing 
$J\gamma^4\,\epsilon = i\alpha\,\epsilon$, which yields
\eq
  \left( \begin{array}{c}
\epsilon_1 \\
\epsilon_2\\
\epsilon_3\\
\epsilon_4 
\end{array} \right) \;\;=\;\;   \left( \begin{array}{c}
\epsilon_1 \\
\epsilon_2\\
-i\alpha\epsilon_1\\
-i\alpha\epsilon_2
\end{array} \right) \,,
\en
where $\alpha=\pm$.

Equations $\delta\lambda, \delta{\psi}_t$ and $\delta\psi_1$
are actually equivalent to each other and can be written in the form
\eq
\delta\lambda\propto \del_\mu S\,\gamma^\mu\,\epsilon
\en
with $S=-\Psi + ie^{-2\phi}$, provided that $\alpha=+1$. 

The equation $\delta \lambda = 0$
can be solved by assuming that $S$ is a holomorphic function
of $z$, therefore $\quad \del_{\bar{z}}S=0$. Note that a holomorphic
function $S(z)$ solves the equation of motion (\ref{phi}).

Then, $\delta\lambda\propto\del_zS\,\gamma^z\,\epsilon=0,$ which can be solved
by setting $\;\;\gamma^z\,\epsilon=0$.

By using  that $\;\gamma^z\propto\gamma^1(1+i)-\gamma^2(1-i),\;$ one finds
\eq
  \left( \begin{array}{c}
\epsilon_1 \\
\epsilon_2\\
\epsilon_3\\
\epsilon_4 
\end{array} \right) \;\;=\;\; \tilde{\epsilon}  \left( \begin{array}{c}
1 \\
i\\
-i\\
1
\end{array} \right) \,.
\en

The remaining equations to be solved are then $\delta{\psi}_r=0$ and
$\delta{\psi}_\theta =0. $ We find
\eqn
&&\del_r\,\tilde{\epsilon}+\hhalf{{e^{2\phi}}\over
  r}\del_\theta\Psi\,\tilde{\epsilon }=0\;\;,\no\\
&&\del_\theta\,\tilde{\epsilon}-{i\over 2}\tilde{\epsilon}-{r\over
  4}e^{2\phi}\del_r\Psi\,\tilde{\epsilon}=0.
\enq
The  holomorphicity of $S$ implies the relations 
\eqn
\del_r\Psi&=&{2\over r}e^{-2\phi}\,\del_\theta\phi\;\;,\no\\
{1\over r}\,\del_\theta\Psi&=&-2\del_r\phi\, e^{-2\phi}\;\;,
\enq
which we use to solve for the spinors completely: 
\eq
\tilde{\epsilon}\,=\,e^{{\phi}\over 2}\,e^{{i\over 2}\theta}.
\en
The above solution breaks 1/2 of $N=8, D=3$ supersymmetry.

\section{String solutions with $H_{\mu\nu\rho}\neq 0$ \label{hmu}}

\setcounter{equation}{0}

Here we show how the Killing spinor equations (\ref{killing}) 
determine a  three-dimensional solution with a non-vanishing
$H_{\mu\nu\rho}$ and a non-constant dilaton (other solutions with non-zero
$H_{\mu\nu\rho}$ have been considered in \cite{hor,hor2}).

We consider a solution without  gauge fields and with constant internal
metric $G_{mn}$. 
We will take $H_{\mu\nu\rho}=\sqrt{-g}\epsilon_{\mu\nu\rho}\Lambda e^{p\phi}$,
where $p$ is an integer, in the Einstein frame.
We will take the space--time metric to be of the form (\ref{gg}), and
we will also take $\phi=\phi(r)$.

The Killing spinor equations reduce to
\eqn
\delta \chi^I&=&0\;\;,\no\\
\delta\lambda &=& -\half e^{-\phi}\del_\mu (\phi+\ln\det e_m^a)
\gamma^\mu\,\varepsilon +{1\over {12}}e^{-3\phi}H_{\mu\nu\rho}\,
\gamma^{\mu\nu\rho}\,\varepsilon\;\;, \no\\
\delta{\psi}_\mu &=&\del_\mu\varepsilon + {1\over 4}\omega_{\mu\alpha\beta}
\gamma^{\alpha\beta}\varepsilon 
 +\hhalf (e_{\mu\alpha}e_\beta^\nu -e_{\mu\beta}e_\alpha^\nu)
\del_\nu\phi\gamma^{\alpha\beta}\varepsilon 
 -{1\over 8}e^{-2\phi}H_{\mu\nu\delta}\gamma^{\nu\delta}\varepsilon\;\;,\no\\
\delta\psi_d &=&-{1\over 4}e^{-\phi}(e_d^m\del_\mu e_{ma}\!+\!e_a^m\del_\mu 
e_{md})\gamma^\mu\gamma^4\otimes\Sigma^a\varepsilon\;\;.
\enq
By using identity (\ref{eps}) of the appendix, these equations become
\eqn
\delta \chi^I&=&0\;\;,\no\\
\delta\lambda &=& -\half e^{-\phi}\del_r (\phi+\ln\det e_m^a)\sqrt{V}
\gamma^1\,\varepsilon +\half e^{(p-3)\phi}\Lambda\,J\,\varepsilon \;\;,\no\\
\delta{\psi}_t&=&\half [\sqrt{V}\del_r\sqrt{V}+V
\del_r\phi]\,J\,\gamma^2\,\varepsilon +{1\over 4} e^{(p-2)\phi}
\sqrt{V} \Lambda\,J\gamma^0\,\varepsilon\;\;,\no\\
\delta {\psi}_r&=&\del_r\varepsilon 
-\hhalf e^{(p-2)\phi}{\Lambda\over{\sqrt{V}}}\,J\gamma^1\,\varepsilon\;\;,\no\\
\delta{\psi}_\theta &=&\del_{\theta}\varepsilon -\half [\sqrt{V}
\del_r R +R\sqrt{V}\del_r\phi]\,J\,\gamma^0\,\varepsilon -\hhalf
e^{(p-2)\phi}R\Lambda\,J\gamma^2 \varepsilon\;\;.
\enq
Compatibility of the spinor
$\varepsilon=\epsilon\otimes\chi$ within these equations 
is obtained by demanding
\eqn
\gamma^1\;\epsilon\;&=&\;\alpha\;J\;\epsilon\;\;,\qquad\alpha=\pm\;,\no\\
\gamma^2\;\epsilon\;&=&\;-\alpha\;\gamma^0\;\epsilon\;\;.
\enq
Then, one has
\eqn
\delta\lambda &=& -\half\alpha e^{-\phi}\del_r (\phi+\ln\det e_m^a)
\sqrt{V}J\,\varepsilon +\half e^{(p-3)\phi}\Lambda
\,J\,\varepsilon \;\;,\no\\
\delta{\psi}_t&=&-\half\alpha [\sqrt{V}\del_r\sqrt{V}+V
\del_r\phi]\,J\,\gamma^0\,\varepsilon +{1\over 4} e^{(p-2)\phi}
\sqrt{V} \Lambda\,J\gamma^0\,\varepsilon\;\;,\no\\
\delta {\psi}_r&=&\del_r\varepsilon 
-\hhalf\alpha e^{(p-2)\phi}{\Lambda\over{\sqrt{V}}}\,\varepsilon\;\;,\no\\
\delta{\psi}_\theta &=&\del_{\theta}\varepsilon -\half [\sqrt{V}
\del_r R +R\sqrt{V}\del_r\phi]\,J\,\gamma^0\,\varepsilon +{{\alpha}\over 4}
e^{(p-2)\phi}R\Lambda\,J\gamma^0 \varepsilon\;\;.
\label{kilhn}
\enq
For $\alpha=+1$ the spinor $\epsilon$ is of the
form
\eq
  \left( \begin{array}{c}
\epsilon_1 \\
\epsilon_2\\
\epsilon_3\\
\epsilon_4 
\end{array} \right) \;\;=\;\;   \left( \begin{array}{c}
\epsilon_1 \\
0\\
\epsilon_3\\
0 
\end{array} \right)\;\;=\;\;\tilde{\epsilon}\left( \begin{array}{c}
1 \\
0\\
1\\
0 
\end{array} \right) \,,
\en
(where we have imposed $\epsilon_1=\epsilon_3$ in order to reduce
 the degrees of freedom contained in $\epsilon$ to the degrees of freedom
 contained in a two component spinor, see appendix),
whereas for  $\alpha=-1$ it is
\eq
\left( \begin{array}{c}
\epsilon_1 \\
\epsilon_2\\
\epsilon_3\\
\epsilon_4 
\end{array} \right) \;\;=\;\;   \left( \begin{array}{c}
0 \\
\epsilon_2\\
0\\
\epsilon_4 
\end{array} \right)\;\;=\;\;\tilde{\epsilon}\left( \begin{array}{c}
0 \\
1\\
0\\
1 
\end{array} \right) \,.
\en

Demanding the vanishing of the Killing spinor equations (\ref{kilhn}) and
imposing the condition $\del_\theta \,\epsilon=0$, as well as
$\alpha\Lambda=|\Lambda |$, leaves us 
with the following constraints 
\eqn
\sqrt{V}\del_r\phi&=&e^{(p-2)\phi}|\Lambda |\;\;,\no\\
  \sqrt{V}\del_r\sqrt{V}+V\del_r\phi &=&
 \half\sqrt{V} e^{(p-2)\phi}|\Lambda |\;\;,\no\\
\del_r\,\tilde{\epsilon} &=&\hhalf
{{|\Lambda|}\over{\sqrt{V}}}e^{(p-2)\phi}\tilde{\epsilon}\;\;,\no\\
\sqrt{V}\del_r R +R\sqrt{V}\del_r\phi &=&\half e^{(p-2)\phi}|\Lambda |\;\;.
\enq
These equations can be solved to give
\eqn
V &=& a^2\,e^{-\phi}\;\;,\no\\
R&=&\,b^2\,e^{-\phi/2}\;\;,\no\\
\tilde{\epsilon}&=&e^{{\phi}\over 4}\epsilon_0,\label{cos}
\enq
where $a,b,\epsilon_0$ are integration constants.

For $p\neq {3\over 2}$, the dilaton is given by
\eq
e^{({3\over 2}-p)\phi}=\bigl|{{\Lambda}\over a}({3\over 2}-p)
(r-r_0)\bigr|,
\en
whereas for $p={3\over 2},$
\eq
\del_r\phi={{|\Lambda|}\over a}
\quad\longrightarrow\quad\phi={{|\Lambda|}\over a}(r-r_0).
\en
Note that $\epsilon$ is real and hence the background preserves 1/2 of $N=8,
D=3$ supersymmetry.

We notice here that whatever the value of $p$, there is a solution to the
Killing spinor equations.
On the other hand, the equations of motion are satisfied provided 
$H^{\mu\nu\rho}={{\epsilon^{\mu\nu\rho}}\over {\sqrt{-g}}}\;\Lambda
\;e^{4\phi}$, that  is $p=4$.
Thus, contrary to common experience \cite{boucher,scha,edel}, 
not every solution to
the Killing spinor equations solves the equations of motion.

For $p=4$, the curvature scalar is computed to be $\, {\cal R}={5\over
  2}\Lambda^2e^{4\phi},\,$ which is always positive.

\section{Conclusions}

We have considered in the present work the low-energy effective Lagrangian of
heterotic string theory compactified 
on a seven-torus, and we have constructed a variety of electrically charged
and solitonic backgrounds preserving 
$1/2^m$ of $N=8, D=3$ supersymmetry ($m = 1,2,3$). The construction of
the solutions is done by using the criteria
 of unbroken supersymmetries and solving for the associated Killing spinor 
equations. 
The space--time line elements of the solutions constructed here 
 have the
form (\ref{met}), which differ from the usual line element associated with
conical geometries \cite{dejahooft}. These line elements do not seem to
correspond to small deformations of flat space--time. Thus they seem to
contain some interesting structure which deserves further study.

Further solitonic solutions with diagonal space--time line elements 
can be obtained by applying more
general $O(8,24)$ transformations to the electrically charged 
solutions of section 4 and 5.
It would  also be interesting to consider 
a non-diagonal ansatz for the space--time metric.

We have also found a solution to the Killing spinor equations with
$H_{\mu\nu\rho}\neq 0$ which preserves 1/2 of the $N=8$ supersymmetry.
Furthermore, we 
have shown that the compactified cosmic string solution constructed by Sen
\cite{sen1} satisfies our Killing spinor equations.
   
Most of the solutions presented here are charged with  the
associated gauge field strengths given everywhere by (\ref{F0r}). 
We note,  however, that one should generally expect 
these solitonic solutions to get modified 
by quantum corrections, at least in the strong coupling regime \cite{sen1}.
Recall the fundamental string solution discussed by Sen \cite{sen1} 
that we have  considered   in section \ref{cosmic1}. There we showed 
 that a holomorphic solution $S(z)$ satisfies our Killing
 spinor equations. As Sen points out, the holomorphic function
 $S(z)=\lambda_1+i\lambda_2$ has the behavior $\lambda_2=e^{-2\phi}
\sim -\ln r\;$ for $\;
 r\longrightarrow 0$, whereas at $r\longrightarrow\infty$, this behavior needs
 to be modified in order to make sense. This is achieved by replacing 
$S$ by the $SL(2,\IZ)$ invariant  function $j(S)$, such that 
$j(S(z))\simeq 1/z\;$ for $\; r\longrightarrow \infty.$ 
Therefore the solution $S(z)$ 
should be considered only as an
approximate solution that gets modified 
as the theory enters the strong coupling regime. 
In analogy with the above, we would, for example, expect
 our electrically charged solutions in subsections \ref{2c} and \ref{1c} 
to get modified  at short distances, where the  coupling 
becomes strong.
Similar phenomena have been shown to occur in $N=4, D=3$ supersymmetric
gauge theories \cite{Seiberg}, where the classical moduli space receives
perturbative as well as non-perturbative quantum corrections. 
An extension of these ideas to string theory remains to be explored.

\bigskip
   
{\bf Acknowledgement}  \medskip \newline
We would like to thank Tom\'as Ort{\'i}n for helpful discussions.


\appendix

\section{Appendix}

\setcounter{equation}{0}

In ten dimensions, the $\Gamma^A$ matrices satisfy
\eq
\{\Gamma^A,\Gamma^B\}=2\eta^{AB}\;\;,
\en
where $A,B$ are D=10  tangent space indices and where 
$\eta_{AB}=(-,+,\dots,+)$.
The decomposition of the gamma matrices that is appropriate to the 10=3+7
split is 
\eqn
\Gamma^A\equiv (\Gamma^\alpha,\Gamma^a)=
 (\gamma^\alpha\otimes {\bf I}_8,\gamma^4\otimes \Sigma^a)\;\;,
\enq
where 
\eqn
\{\gamma^\alpha,\gamma^\beta \}&=& 2\eta^{\alpha\beta},\qquad 
\eta^{\alpha\beta}=(-,+,+)\;\;,\no\\
\{\Sigma^a,\Sigma^b\}&=&2\eta^{ab},\qquad
\,\eta^{ab}=(+,+,\dots ,+)\;\;,\no\\
\{\gamma^4,\gamma^\alpha\}&=&0,\quad\qquad\, (\gamma^4)^2={\bf I}_4\;\;.
\enq
$\gamma^\alpha$ and $\Sigma^a$ are the 3D space--time  and 7D internal 
Dirac matrices respectively,
and $\gamma^4$ plays the role of a chirality operator \cite{wetterich} by 
enhancing the three-dimensional space--time spinor to a four component 
spinor (instead of the
usual two component spinor). Note that in order to have
$\{\Gamma^a,\Gamma^\alpha \}=0$ we are forced  to introduce the chirality
operator $\gamma^4$. 
We decompose the 10D spinor into the form
$\varepsilon^{A,i}=\epsilon^A\otimes\chi^i$, where $\epsilon^A$ is a four 
component spinor of SO(1,2) ($A=1,\dots,4$) and $\chi^i$ is a SO(7) spinor,
with $i=1,\dots,8$ indicating the $N=8$ supersymmetries.

We have
\eqn
&&\Gamma^{\alpha\beta}=\gamma^{\alpha\beta}\otimes {\bf I}_8,
\quad\quad \Gamma^{\alpha a}=\gamma^\alpha\gamma^4\otimes\Sigma^{a},
\quad\quad \Gamma^{ab}={\bf I}_4\otimes\Sigma^{ab}\;\;,\no\\
&&\Gamma^{\alpha a b}=\gamma^\alpha \otimes\Sigma^{ab},\qquad
\Gamma^{\alpha \beta a}=\gamma^{\alpha\beta}\gamma^4\otimes \Sigma^a,
\enq
with $\Gamma^{AB\dots C}=\Gamma^{[a}\Gamma^b\dots\Gamma^{C]}$, thus
$\Gamma^{AB}=\half(\Gamma^A\Gamma^B-\Gamma^B\Gamma^A)$.
The 10D chirality operator $\Gamma^{11}$ is given by
$\Gamma^{11} = \gamma^0  \gamma^1  \gamma^2 \gamma^4 \otimes i  
{\bf I}_8$.

    As a representation of the $\gamma $ matrices, we take
\eq
\gamma^0=  \left( \begin{array}{cl}
0 & i\sigma^2 \\
i\sigma^2 & 0 
\end{array} \right) \;,\;\;\;\;\; \gamma^1=  \left( \begin{array}{cl}
0 & \sigma^3 \\
\sigma^3 & 0 
\end{array} \right)\;,\;\;\;\;\; \gamma^2=  \left( \begin{array}{cl}
0 & \sigma^1 \\
\sigma^1 & 0 
\end{array} \right)\;,\;\;\;\;\; \gamma^4=  \left( \begin{array}{cl}
{\bf I}_2 & 0 \\
0 & -{\bf I}_2 
\end{array} \right)\,, \]
with
\[
\sigma^1=  \left( \begin{array}{cl}
0 & 1 \\
1 & 0 
\end{array} \right) \;,\;\;\;\;\; \sigma^2=  \left( \begin{array}{cl}
0 & -i \\
i & 0 
\end{array} \right)\;,\;\;\;\;\; \sigma^3=  \left( \begin{array}{cl}
1 & 0 \\
0 & -1 
\end{array} \right)\;. 
\en

Then
\eq
\gamma^{01}=-J\gamma^2,\qquad \gamma^{02}=J\gamma^1,
\qquad \gamma^{12}=J\gamma^0\;\;,
\en
where 
\eq
J=  \left( \begin{array}{cl}
0 &{\bf I}_2  \\
{\bf I}_2 & 0 
\end{array} \right)\;\;. 
\en
We  have $\gamma^\mu=e_\alpha^\mu\gamma^\alpha,\quad
\Sigma^m=e_a^m\Sigma^a,$ where $e_\alpha^\mu$ and $e_a^m$ are the space--time 
and internal inverse vielbeins respectively.

    The conventions for the Christoffel symbols and the Ricci 
tensor are the following:
\eqn
&&\Gamma_{\mu\nu}^\rho=\half\sum_\sigma g^{\rho\sigma}\{{{\del g_{\nu\sigma}}
\over{\del x^\mu}}+{{\del g_{\mu\sigma}}\over{\del x^\nu}}
-{{\del g_{\mu\nu}}\over{\del x^\sigma}}\}\;\;,\no\\
&&{\cal R}_{\mu\rho}=\sum_\nu {{\cal R}_{\mu\nu\rho}}^\nu
=\sum_\nu{{\del}\over{\del x^\nu}}
\Gamma^\nu_{\mu\rho}-{{\del }\over {\del x^\mu}}(\sum_\nu\Gamma^\nu_{\nu\rho})
+\sum_{\alpha,\nu}(\Gamma^\alpha_{\mu\rho}\Gamma^\nu_{\alpha\nu}
-\Gamma^\alpha_{\nu\rho}\Gamma^\nu_{\alpha\mu})\;\;,\no\\
&&\sum_\nu \Gamma^\nu_{\nu\mu}=\half \sum_{\nu,\alpha}g^{\nu\alpha}
{{\del g_{\nu\alpha}}\over{\del x^\mu}}={{\del}\over{\del x^\mu}}
\ln \sqrt{|g|}\;\;.\label{ricci1}
\enq
Our convention  for the spin connection is
\eq
\omega_{MAB}=\half E_A^N(\del_ME_{NB}-\del_NE_{MB})+\half E_B^N(\del_NE_{MA}
-\del_ME_{NA})-\half E_A^PE_B^Q(\del_PE_Q^C-\del_QE_P^C)E_{MC}
\en
    and 
\eq
G_{MN}=E_M^A\,\eta_{AB}\,E_N^B, \qquad\qquad E_{MB}=E_M^C\,\eta_{CB}.
\en
Other useful identities are
\eqn
\epsilon_{\mu\nu\rho}\;\gamma^{\mu\nu\rho}&=&
{1\over {\det e}}\;\epsilon_{\alpha\beta\eta}\;\gamma^{\alpha\beta\eta}\no\\
&&={6\over {\det e}}\;\epsilon_{012}\;\gamma^{012}
\qquad\qquad {\rm with }\;\epsilon^{012}=-\epsilon_{012}=1\;\;,\no\\
\epsilon_{\mu\nu\rho}\;\gamma^{\nu\rho}&=&{1\over {\det e}}\;e_\mu^{\alpha}\;
\epsilon_{\alpha\beta\eta}\;\gamma^{\beta\eta}\;\;.\label{eps}
\enq

\goodbreak

\bigskip  \bigskip


\end{document}